\documentclass[journal,draftclsnofoot,onecolumn,12pt]{IEEEtran}

\usepackage{filecontents}
\usepackage{glossaries}
\usepackage[table,xcdraw]{xcolor}
\usepackage{amsthm}
\usepackage{epsfig}
\usepackage{times}
\usepackage{verbatim}
\usepackage{enumerate}
\usepackage{multicol}
\usepackage{afterpage}
\usepackage{wrapfig}
\usepackage{cite}
\usepackage{graphicx}
\usepackage{algorithm}
\usepackage{algpseudocode}
\usepackage{ifmtarg}
\usepackage{amssymb}
\usepackage{amsmath}
\usepackage{color}
\usepackage{bbm}
\usepackage{bm}
\usepackage{epstopdf}
\usepackage{flushend}
\usepackage{xspace}
\usepackage[caption=false,font=footnotesize]{subfig}
 \usepackage{multirow}
 \usepackage{stackengine}

\newtheorem{theorem}{Theorem}
\newtheorem{corollary}{Corollary}

\newtheorem{lemma}{Lemma}

\theoremstyle{remark}
\newtheorem{remark}{    Remark}
\def\blam{{\boldsymbol{\lambda}}}
\def\bmu{{\boldsymbol{\mu}}}
 
\def\bb0{{\mathbb{0}}}

\def\bbE{{\mathbb{E}}}

\def\bbP{{\mathbb{P}}}

\def\ba{{\mathbf{a}}}
\def\bb{{\mathbf{b}}}

\def\bff{{\mathbf{f}}}

\def\bh{{\mathbf{h}}}

\def\bw{{\mathbf{w}}}

\def\b0{{\mathbf{0}}}

\def\bA{{\mathbf{A}}}

\def\bC{{\mathbf{C}}}

\def\bF{{\mathbf{F}}}
\def\bG{{\mathbf{G}}}
\def\bH{{\mathbf{H}}}
\def\bI{{\mathbf{I}}}

\def\cA{\mathcal{A}}

\def\cC{\mathcal{C}}

\def\cE{\mathcal{E}}

\def\cM{\mathcal{M}}
\def\cN{\mathcal{N}}

\def\cT{\mathcal{T}}

\def\cV{\mathcal{V}}

\def\sfR{\mathsf{R}}

\def\sf0{{\mathsf{0}}}

\newcommand{\figref}[1]{Fig.~\ref{#1}}

\newcommand{\thmref}[1]{Theorem~\ref{#1}}
\newcommand{\lemref}[1]{Lemma~\ref{#1}}
\newcommand{\corref}[1]{Corollory~\ref{#1}}
\newcommand{\secref}[1]{Section~\ref{#1}}

\newcommand{\appref}[1]{Appendix~\ref{#1}}

\newcommand{\parens}[1]{\left(#1\right)}

\newcommand{\bars}[1]{\left\vert#1\right\vert}

\newcommand{\floor}[1]{\left\lfloor{#1}\right\rfloor}

\newcommand{\msnr}{\mathrm{SNR}}
\newcommand{\msinr}{\mathrm{SINR}}
\newcommand{\mrinr}{\mathrm{RINR}}
\newcommand{\minlam}{\lambda_{\min}}
\newcommand{\hdrm}{\mathrm{HD}}
\newcommand{\fdrm}{\mathrm{FD}}
\newcommand{\rxrm}{\mathrm{rx}}
\newcommand{\txrm}{\mathrm{tx}}

\newcommand{\aoa}{\textrm{AoA}}
\newcommand{\aod}{\textrm{AoD}}
\newcommand{\targetdelta}{\delta_{\mathrm{target}}}
\newcommand{\maxrm}{\max}
\newcommand{\ttar}{\zeta}
\newcommand{\nrays}{N_{\textrm{ray}}}
\newcommand{\nclusters}{N_{\textrm{cluster}}}
\newcommand{\nbs}{N_{\textrm{BS}}}

\newcommand{\indicator}[1]{\mathbbm{1}_{#1}}

\newcommand{\snr}{\gls{snr}\xspace}
\newcommand{\sinr}{\gls{sinr}\xspace}

\newcommand{\rinr}{\gls{rinr}\xspace}

\newcommand{\mmw}{\gls{mmw}\xspace}
\newcommand{\ula}{\gls{ula}\xspace}

\newcommand{\dft}{\gls{dft}\xspace}

\newcommand{\iab}{\gls{iab}\xspace}

\newcommand{\hd}{\gls{hd}\xspace}
\newcommand{\fd}{\gls{fd}\xspace}
\newcommand{\si}{self-interference\xspace}
\newcommand{\cdf}{\gls{cdf}\xspace}

\newcommand{\st}{\mathrm{subject~to}}

\newacronym{snr}{SNR}{signal-to-noise ratio}
\newacronym{sinr}{SINR}{signal-to-interference-plus-noise ratio}
\newacronym{inr}{INR}{interference-to-noise ratio}
\newacronym{rinr}{RINR}{residual-interference-to-noise ratio}
\newacronym{los}{LOS}{line-of-sight}
\newacronym{nlos}{NLOS}{non-line-of-sight}
\newacronym{cdf}{CDF}{cumulative density function}
\newacronym{pdf}{PDF}{probability density function}
\newacronym{fd}{FD}{full-duplex}
\newacronym{hd}{HD}{half-duplex}
\newacronym{si}{SI}{self-interference}
\newacronym{mmw}{mmWave}{millimeter wave}
\newacronym{fr2}{FR2}{frequency range 2}
\newacronym{iab}{IAB}{integrated access and backhaul}
\newacronym{ula}{ULA}{uniform linear array}
\newacronym{rf}{RF}{radio frequency}
\newacronym{dft}{DFT}{discrete fourier transform}

\newcommand{\mycomment}[1]{}
\usepackage{mathtools} 

\makeatother
\begin{document}
\title{System-Level Analysis of Full-Duplex Self-Backhauled Millimeter Wave Networks}


	\author{
		\IEEEauthorblockN{\large  Manan Gupta,  Ian P. Roberts, and Jeffrey G. Andrews }\\
		\thanks{The authors are with the Wireless Networking and Communications Group (WNCG), The University of Texas at Austin, Austin, TX. Email: \{g.manan@utexas.edu, ipr@utexas.edu, jandrews@ece.utexas.edu\}. Last modified: \today. }}
	
	\maketitle
	\begin{abstract}
	    \Gls{iab} facilitates cost-effective deployment of \mmw cellular networks through multihop self-backhauling.
	    \Gls{fd} technology, particularly for \mmw systems, is a potential means to overcome latency and throughput challenges faced by \iab networks.
	    We derive practical and tractable throughput and latency constraints using queueing theory and formulate a network utility maximization problem to evaluate both \fd-\iab and \hd-\iab networks.
	    We use this to characterize the network-level improvements seen when upgrading from conventional \hd \iab nodes to \fd ones by deriving closed-form expressions for (i) latency gain of \fd-\iab over \hd-\iab and (ii) the maximum number of hops that a \hd- and \fd-\iab network can support while satisfying latency and throughput targets.
	    Extensive simulations illustrate that \fd-\iab can facilitate reduced latency, higher throughput, deeper networks, and fairer service.
	    Compared to \hd-\iab, \fd-\iab can improve throughput by $8\times$ and reduce latency by $4\times$ for a fourth-hop user.
	    In fact, upgrading \iab nodes with \fd capability can allow the network to support latency and throughput targets that its \hd counterpart fundamentally cannot meet. The gains are more profound for users further from the donor and can be achieved even when residual self-interference is significantly above the noise floor.
	\end{abstract}

\glsresetall

\section{Introduction}
\label{Sec:Intro}

Noteworthy hurdles exist in the cost-effective deployment of \mmw cellular networks that can reliably supply users with high data rates and low latency---stemming largely from the severe pathloss and blockage vulnerability when communicating at such high carrier frequencies \cite{Pi11,Millimeter_Rappaport13, heath16overview, rangan2014millimeter}.
\Gls{iab} is a promising means to deploy \mmw networks with the base station (BS) density necessary to deliver reliable, widespread coverage 
\cite{3gpp_tr38874,cudak21integrated,rasek20joint,gupta2020andrews,ortiz19scaros,du17gbps,nazmul17sampath,kulkarni17performance}.
\iab is a multihop network deployment where the majority of BSs---called \textit{\iab nodes}---wirelessly backhaul their traffic to fiber-backhauled \textit{donor nodes}, possibly relaying through other \iab nodes. 

This architecture attractively offers a reduction in the number of fiber connections needed to deploy a \mmw network, making dense networks practically viable. 
In return, however, packets are relayed through the network, leading to poor rate scaling and packet delays, which is particularly concerning for delay-sensitive applications like video calls, gaming, and virtual reality.
To meet the strict quality-of-service requirements for these modern applications, \fd technology provides an approach to address resource bottlenecks---potentially improving rate scaling and latency---and augments existing resource allocation solutions.
Equipping \iab nodes with \fd capability allows them to simultaneously transmit and receive over the same bandwidth, virtually doubling the available radio resources compared to conventional \hd operation.
It is well known that this transceiver-level upgrade can directly translate to link-level gains, but the network-level gains are less clear.
In this paper, we study the potential gains in network performance---in terms of throughput, latency, and network depth---when upgrading from \hd \iab nodes to ones with \fd capability, which may transcend the potential doubling of spectral efficiency.


\subsection{Motivation, Background, and Related Work}

Multihop networks have been an active area of research for a few decades \cite{jain2005impact, gupta00capacity, fra09capaccity}, as they require fewer fiber connections to tessellate an area according to the given coverage criterion, for example, received \snr above a threshold .
However, per-user throughput deteriorates and packet delays increase with the number of hops between the donor and a user-equipment (UE) \cite{gupta00capacity,zemlianov05capacity}.
Even though \iab networks operating at \mmw frequencies benefit from larger bandwidth (offering high data rates on the backhaul) and reduced interference (from directional communication and raised integrated noise power), they are subject to the same fundamental \textit{throughput-coverage} trade-off as conventional multihop networks. 
As the number of hops increase, the throughput and latency performance degrade due to packet relaying, buffering, and link multiplexing delays \cite{cudak21integrated, polese2020integrated}.

As a result, to satisfy throughput and latency targets, efficient resource utilization at the \iab node is of utmost importance and has led to a variety of studies on route selection \cite{ortiz19scaros, polese2018distributed, vu19joint}, link scheduling \cite{cuba2020twice, gupta2020andrews}, load balancing \cite{saha19millimeter}, and topology optimization \cite{madapatha2021topology,nazmul17sampath}.
In \cite{ortiz19scaros, gupta2020andrews}, authors propose reinforcement learning frameworks that aim to minimize end-to-end latency of packets, and \cite{polese2018distributed} presents routing strategies that minimize the number of hops to improve throughput and latency. 
Link scheduling and power allocation solutions that leverage simulated-annealing are proposed in \cite{cuba2020twice}, the benefits of offloading UEs to \iab nodes is studied in \cite{saha19millimeter}, and genetic algorithm based schemes for \iab node placement and non-\iab link distribution are developed in \cite{madapatha2021topology}.
For a comprehensive survey on recent developments in \iab, please see \cite{zhang2021survey}. 
In this work however, we investigate how \fd capability can alleviate the aforementioned resource bottleneck and latency issues present in \iab networks.
Recent breakthroughs in \si cancellation using analog \cite{roberts20equipping}, digital \cite{bishnu21performance}, and spatial \cite{roberts21millimeter, xiao17fullduplex} cancellation techniques can rid a receive signal of \si. 
In \cite{suk21fullduplex}, the authors prototype a two-hop \iab network and show that the throughput for \fd-\iab is almost twice that of \hd-\iab. 
In \cite{bishnu21performance}, the authors evaluate a multi-user \fd-\iab network and present user selection and digital \si cancellation techniques to maximize the received signal power at the user. 
These works, however, do not explore the network-level consequences of \mmw \fd in an \iab deployment.
In this paper, we aim to quantify the relative gain of \fd-\iab over \hd-\iab in terms of the UE throughput, network depth, and latency in realistic multihop \iab deployments.

\subsection{Contributions}

Our technical contributions are summarized as follows.

\textbf{A single optimization framework for \hd-\iab and \fd-\iab networks.}
We formulate an optimization problem to study \textit{latency} and \textit{throughput}---arguably the most important performance metrics---in a \mmw \iab network. 
In \secref{Sec:SysMod}, we model the \iab network as a Jackson network of queues, which allows us to leverage results from queueing theory to model packet delay as the sojourn time and throughput as the packet arrival rate. 
We use this queueing model to formulate a utility maximization with throughput and latency constraints that are both practical and tractable. 
Our network optimization problem is a convex program and is parameterized by the routing structure of the network, the link capacities, and scheduling restrictions on each BS, which in turn depend on whether the BS is capable of \hd or \fd communication.
Solving the convex program returns the per-user throughput and the corresponding resource allocation scheme that maximizes a desired network utility.

\textbf{Characterizing the network gain of \fd-\iab over \hd-\iab.}
In \secref{Sec:PerAnalysis}, we use the optimization framework to compare \fd-\iab with \hd-\iab and derive a closed-form expression for \textit{latency gain}. 
We also derive the maximum number of hops that achieve a target latency and target throughput for each UE, for both \fd-\iab and \hd-\iab. 
In \secref{subsec:rate_latency_gain}, we show numerical results for both \textit{rate gain} and latency gain of \fd-\iab over \hd-\iab.
The benefit of equipping the \iab nodes with \fd transceivers is more subtle and powerful than the familiar link-level gains (i.e., what is typically less than a two-fold improvement in spectral efficiency).
We show that the more hops between a UE and the donor, the more the UE has to gain from a \fd-\iab deployment, both in terms of latency and throughput.
For example, the throughput of a UE four hops from the donor can improve by $8\times$ if \iab nodes are upgraded to \fd.
This many-fold increase stems from the increased scheduling opportunities that FD provides at each \iab node, which reduces multiplexing delays that are particularly significant in the context of multihop routing.

\textbf{Impact of imperfect \si cancellation on network performance.}
In \secref{subsec:rate_latency_gain} we present numerical results to quantify the effect of imperfect \si cancellation on the throughput performance of a \fd-\iab network.
We show that even if the residual \si is $10$ dB above the noise floor, a six-fold rate improvement can be achieved by \fd-\iab over its \hd counterpart.
Reducing the \rinr below $0$ dB can result in $8$ times rate gain for a user at the fourth hop.
The \fd gains saturate for \rinr below $-5$ dB, suggesting that---from a network perspective---further self-interference cancellation is likely not worthwhile.
These insights can drive physical layer design decisions for \fd in \iab networks.

\textit{Notation}: $\bA_{:,i}$ denotes the $i$-th column of matrix $\bA$, $\bA_{i,:}$ denotes the $i$-th row of matrix $\bA$, $\bA^T$ denotes the transpose of a matrix, $[\bA]_{i,j}$ denotes the $(i,j)$-th element of $\bA$, and $\bA^\dagger$ denotes the hermitian transpose of a matrix. The $i$-th element of vector $\ba$ is denoted by $\ba_i$. $\indicator{\cA}$ denotes the indicator function over set $\cA$, and $|\cA|$ denotes the cardinality of set $\cA$.
\section{System Model}
\label{Sec:SysMod}

\begin{figure}[t!]
    \centering
    \includegraphics[width=4.5in,                 keepaspectratio]{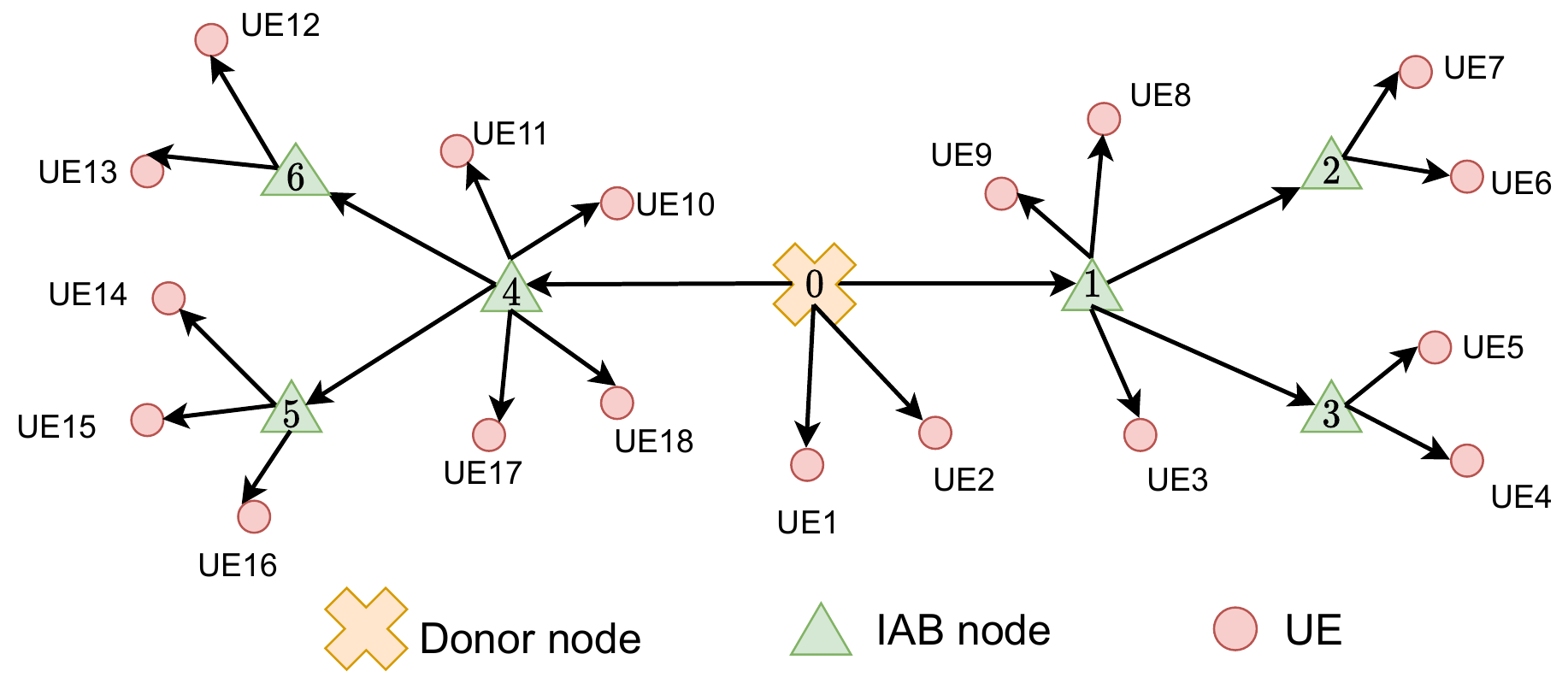}
    \caption{A multihop routing tree with one donor node, $K=6$ \iab nodes, and $M=18$ UEs. Note that each \iab node has one parent but can serve multiple children.}
    \label{fig:IABtree}
\end{figure}

Illustrated in \figref{fig:IABtree}, we consider a downlink \iab deployment with one fiber-backhauled donor node and $K$ \iab nodes which wirelessly backhaul to the donor, possibly through multiple hops. 
$M$ UEs are present in the network, each of which can either be served by the donor or an \iab node. 
Downlink data arrives at the donor from the network core and is delivered to the UEs, using the \iab nodes as relays as needed.
In this work, we will consider and evaluate both \hd- and \fd-equipped \iab nodes.
We assume UEs are conventional \hd devices.
We represent the flow of data---the route between the source (donor) and a destination (UE)---through a \textit{routing tree}, $\cT = (\cV, \cE)$, where $\cV$ denotes the set of vertices (devices) such that $|\cV| = K+M+1$ and includes the \mmw BSs and the UEs. 
Henceforth, the term \textit{device} will be used to refer to the donor node, an \iab node, or a UE.
$\cE$ denotes the set of edges in the tree $\cT$ such that if $(u, v)\in\cE$ then $u, v \in \cV$ and there exists a directed edge with $u$ as the parent and $v$ as the child. 
In other words, a device $v$ receives downlink data from a device $u$. 
Since all vertices, except for the donor, have one (and only one) parent edge, we have $|\cE| = K+M$. 
Since an edge $(u,v)$ is uniquely identified by its child $v$, we will index edge $(u,v)$ as $v$. 
An illustration of an example \iab routing tree is shown in \figref{fig:IABtree}, where vertex $0$ represents the donor and triangles denote \iab nodes. 
IAB1 (vertex $1$) is one hop away from the donor whereas IAB2 (vertex $2$) is two hops away since it communicates with the donor via IAB1. 
Similarly, UE1, UE3, and UE5 are one-hop, two-hop, and three-hop UEs, respectively.

\textit{Link Capacities and Scheduling:}
Associated to each edge $(u,v) \in\cE$ is the capacity of the edge $c_v$ based on its link quality; note that $c_v$ is only indexed by $v$ since each edge $(u,v)$ is uniquely identified by its child (i.e., each device has only one parent).
In an \iab network, simultaneously transmitting data on all edges in $\cT$ is not possible due various hardware and design constraints, such as \hd constraints and/or limited multi-user communication capabilities.
A network scheduler typically decides the set of vertices that can communicate at a given time, forming the set of \textit{active} edges.
Thus, each edge is only allocated a certain fraction of the time by the scheduler to transmit data. 
We denote by $\mu_v$ the fraction of time allocated to edge $(u,v)$ for data transmission, meaning $c_v\mu_v$ represents the effective long-term data rate of the edge $(u,v)$. 



\textit{Modeling the \iab Network as Network of Queues:}
In order to analyze the delay distribution of UEs across different hops, we model the \iab network as a network of queues. 
Data for each UE arrives in packets at the IAB donor following a stochastic process and must be delivered to the destined UE along the route given by $\cT$. 
We denote by $A_m(t)$ the number of packets destined for the $m$-th UE that arrives at the donor at time $t$, where $\lambda_m = \bbE[A_m(t)]$
denotes the mean of the arrival stochastic process for packets intended for the $m$-th UE (i.e., the \textit{arrival rate}).
We use $\blam$ to denote the $M\times 1$ vector of mean arrival rates $\lambda_m$ for the $M$ UEs in the network. 
Each edge $(u,v)\in\cT$ maintains a queue to buffer packets that $u$ must transmit to $v$ and is referred to as queue $(u,v)$. 

\begin{figure}
    \centering
    \subfloat[Multihop routing tree.]{%
        \includegraphics[width=\linewidth,height=0.13\textheight,
                        keepaspectratio]{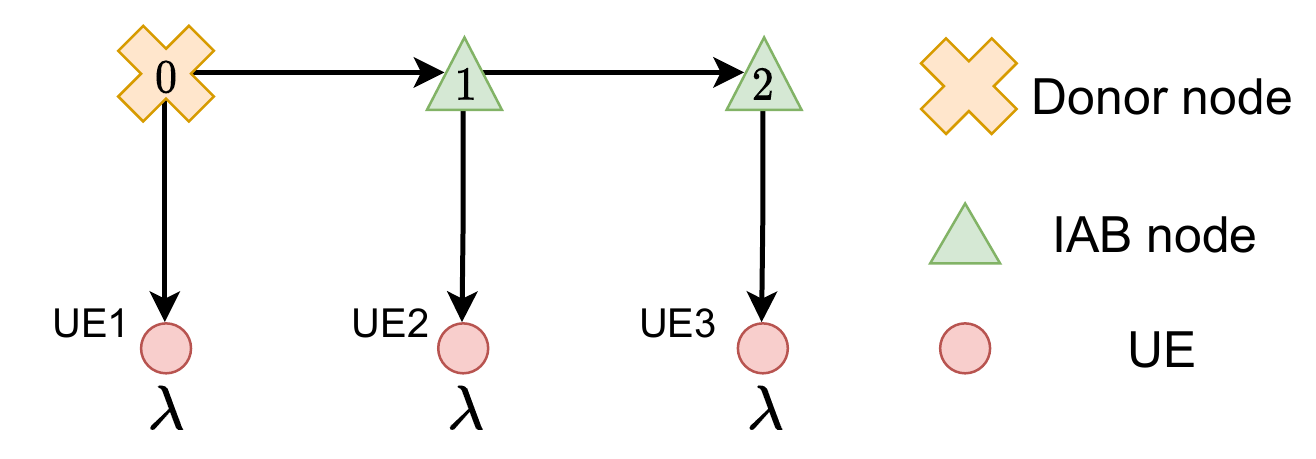}%
        \label{fig:3HopNet}}\\
    \subfloat[Equivalent queueing network.]{%
        \includegraphics[width=\linewidth,height=0.13\textheight,keepaspectratio]{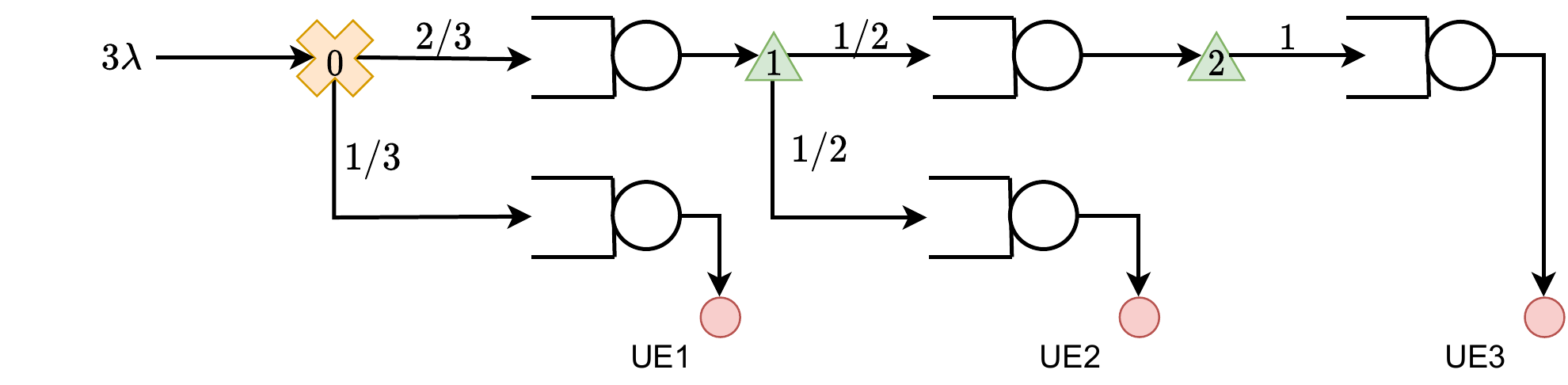}%
        \label{fig:3HopNetJackson}}
    \caption{(a) A multihop routing tree with $K=2$ \iab nodes and $M=3$ UEs. The arrival rate of packets intended for each UE is $\lambda$. (b) The equivalent queueing network of (a) with traffic splitting at the donor and each \iab node. The total arrival rate into the network is $3\lambda$.}
    \label{fig:Tree2Queue}
\end{figure}

Let $\bF$ denote the $|\cE|\times M$ routing matrix such that $[\bF]_{l,m} = 1$ if traffic for the $m$-th UE is routed through the $l$-th edge and is zero otherwise.
The arrival rate to each queue is given by the vector $\bF\blam$. 
Since the effective long-term data rate of edge $(u,v)$ is $c_v\mu_v$, the mean service time for queue $(u,v)$ is given by $1/(c_v\mu_v)$.
Modeling each edge as a queue equivalently implies that the donor and \iab nodes maintain queues to buffer packets for each of their children. 
A packet arriving at BS $k$ is placed in queue $(k,v)$ with probability $(\bF\blam)_{v} / (\bF\blam)_{k}$, as shown in \figref{fig:3HopNetJackson}. 
This ensures that the average number of packets delivered to each UE is the same as the number of packets arriving at the donor destined for that UE.

\section{Formulating an IAB Network Design Problem}
\label{Sec:ProbForm}
With an \iab network model in place, we will now formulate a network utility maximization subject to practical delay and throughput constraints. 
Our goal is to use this optimization problem to study the network performance improvement---in terms of throughput and latency---when the \hd transceivers at the \iab nodes are upgraded to \fd ones, for a given deployment.
Our optimization problem is parameterized by the network topology and the choice of either \hd- or \fd-equipped \iab nodes. 
Specifically, we use a routing matrix $\bF$ to describe the routes between the donor and the UEs and is readily obtained for a given routing tree.
We use a capacity matrix $\bC$ and a scheduling matrix $\bG$ to describe the effective long-term rate of each link, both of which depend on the choice of \hd or \fd \iab nodes. 
Feasibility constraints enforce that each link meets its demand. 
A latency constraint ensures that a fraction of packets $\eta$ are delivered to their target UE within $\delta$ units of time. 
We aim to find the average arrival rates $\blam$ and resource allocations $\bmu$ that maximize a chosen network utility.
We outline these components of our design in detail as follows and subsequently assemble our network design optimization problem.

\textbf{Constraint 1: Fundamental constraints on arrival rate and resource fractions.}
As follows, we capture the fundamental properties of the arrival rates $\blam$, each of which must be non-negative, and resource allocations $\bmu$, each of which must lie in the interval $[0,1]$.
\begin{align}
    \blam &\geq \bm{0}, \quad \bm{0} \leq \bmu \leq \bm{1}
\end{align}

\textbf{Constraint 2: Constraint on feasible arrival rate.}
The \sinr of the signal at device $v$ when receiving from its parent $u$ is given by $\msinr_{v}$, which depends on a variety of environmental and system factors such as pathloss and transmit/receive beamforming. 
The capacity of the edge $(u,v)$ is
\begin{align}
c_v = W \cdot \log_2(1 + \msinr_{v})
\end{align}
where $W$ is the system bandwidth. 
Note that \fd \iab nodes incur \si, making \sinr the metric of interest. 
\hd \iab nodes are effectively interference-free under noise-limited network conditions, meaning we can simply use the \snr.

Often, \mmw communication leads to noise-limited conditions due to highly directional communication, severe pathloss at \mmw frequencies, and high susceptibility to blockages  \cite{Millimeter_Rappaport13,bai2015coverage}.
The severity of network interference in \fd-\iab network is still an open question, but for tractability we assume it is negligible, largely thanks to the aforementioned properties of \mmw communication \cite{rangan2014millimeter, yuan2018optimal}.
Furthermore, the donor and other \iab nodes would likely be pole-mounted and presumably deployed to avoid directed interference on the backhaul transmissions, while UEs are likely on the ground.
Together, these establish sufficient spatial isolation between transmitting and receiving devices that happen to be nearby and scheduled simultaneously.
Also, since the UEs are \hd, backhaul transmissions would not suffer interference from arbitrary directions, and as observed in \cite{kulkarni19maxmin}, a competent scheduling algorithm further reinforces the noise-limited behavior since it avoids scheduling interfering links in the same slot.
Our simulations also suggest noise-limited behavior for both \hd- and \fd-\iab deployments.

We denote by $\bC$ the $|\cE|\times |\cE|$ diagonal matrix for edge capacities $c_v$.
Then, the product $\bC\bmu$ is the vector containing the long-term average rate of each edge or, equivalently, the vector of mean service rates of each queue. 
For the queues to be stable, the mean service rate of each queue should be greater than its mean traffic arrival rate, leading to
\begin{align}
\bC\bmu > \bF\blam.
\end{align}

\textbf{Constraint 3: Scheduling constraints.}
We denote by $\bG$ the $(K+1)\times |\cE|$ scheduling matrix for the network, which defines the resource constraints on each BS (donor node or \iab node). 
More precisely, $[\bG]_{k,v} = 1$ if BS $k$ must allocate orthogonal time resources to edge $(k,v)$, and zero otherwise.
For example, for the routing tree in \figref{fig:3HopNet} the \hd scheduling matrix $\bG_{\hdrm}$ and the \fd scheduling matrix $\bG_{\fdrm}$ are given in \eqref{eq:scheduling_mat_ex}.
For both matrices, the second row represents IAB1. In the \hd case, it must split resources between receiving from the donor, transmitting to UE1, and transmitting to IAB2. 
On the other hand, when IAB1 is \fd-capable, it can receive while transmitting, and must no longer dedicate resources to edge $(0,1)$.
\begin{align}
    \label{eq:scheduling_mat_ex}
    \bG_{\hdrm} = \begin{bmatrix} 
        1 & 1 & 0 & 0 & 0 \\
        0 & 1 & 1 & 1 & 0 \\
        0 & 0 & 0 & 1 & 1
    \end{bmatrix}\quad \bG_{\fdrm} = \begin{bmatrix} 
        1 & 1 & 0 & 0 & 0 \\
        0 & 0 & 1 & 1 & 0 \\
        0 & 0 & 0 & 0 & 1
    \end{bmatrix}
\end{align}

\textbf{Constraint 4: Probabilistic latency constraint.}
If the arrival processes $A_m(t)$ are Poisson and the queues have exponential service times, then the queueing network in \figref{fig:3HopNetJackson} is a \textit{Jackson Network} \cite[Section 9.9.1]{srikant2014comnets} without feedback loops, and the steady-state joint distribution of the queueing network is product-form.
Both the Poisson arrival process and the exponentially varying packet size \cite{3gpp_tr36889} are well-accepted models.
As a result, we can treat each queue $(u,v)$ as an independent $M/M/1$ queue with input arrival rate $(\bF\blam)_v$ and mean service rate $c_v\mu_v$ (the effective capacity from device $u$ to device $v$). 
Consequently, if $D_v$ denotes the random variable representing the delay experienced by a typical packet over queue $(u,v)$, then $D_v$ follows an exponential distribution with the following \cdf.
\begin{align}
    \label{eq:cdf_delay}
    \bbP[D_v\leq d] = 1 - \mathrm{exp}\parens{-(c_v\mu_v - (\bF\blam)_{v}) \cdot d}
\end{align} 

Let $\sfR(s,d) = ((s,u_1), (u_1, u_2), \dots, (u_k, d))$ denote the route between source-destination pair $(s,d)$. A route is defined as the sequence of edges in $\cT$ traversed by a packet to go from $s$ to $d$. For brevity, we will denote by $\sfR(m)$ the route between the donor and UE $m$. 
Let $D_m$ be defined as the total delay experienced by a packet destined for the $m$-th UE, which is simply the sum of delays the packet incurred along its route.
\begin{align}
    D_m = \sum_{(u,v)\in \sfR(m)}D_v
\end{align}
Since the per-hop delays $\{D_v\}$ are independent exponential random variables with different means, the sum delay $D_m$ follows a hypoexponential distribution which is non-convex and intractable.
Thus, for our analysis we will use a stricter notion of delay.
Let $\tilde{D}_m$ denote the maximum delay experienced on any hop in its route to UE $m$. 
\begin{align}
    \tilde{D}_m = \max_{(u,v)\in \sfR(m)} D_v
\end{align}
It trivially holds that $D_m \leq h_m\tilde{D}_m$, where $h_m = \sum_{(u,v)}\indicator{\{(u,v)\in \sfR(m)\}}$ denotes the number of hops between the donor and UE $m$. The CDF of $\tilde{D}_m$ is given by 
\begin{align}
\label{eq:max_delay_CDF}
    \bbP\left[h_m\tilde{D}_m \leq d\right] 
                                                &=\prod_{(u,v)\in \sfR(s,d)}\left(1 - e^{-(c_v\mu_v - (\bF\blam)_{v})d/h_m}\right).
\end{align}
If we desire that a fraction $\eta$ of packets are delivered to their target UE within $\delta$ units of time, we can form a probabilistic latency constraint as $\bbP[h_m\tilde{D}_m\leq \delta] > \eta$, or equivalently $\log(\bbP[h_m\tilde{D}_m\leq \delta]) > \log(\eta)$. Mathetically, we have
\begin{align}
    \sum_{(u,v)\in \sfR(m)}\log\left(1 - e^{-(c_{u,v}\mu_{u,v} - (\bF\blam)_v)\delta/h_m}\right) &\geq \log(\eta), \quad m = 1, 2, \dots, M.
\end{align}
Note that $\delta$ is the parameter which reflects the delay threshold of the network, whereas $\eta$ sets how strict it is that the threshold is met.


\textbf{Objective: Maximize network utility.}
As is widely used by network designers, the objective of our design is to maximize sum utility of the network. 
We denote by $U(\cdot)$ a non-decreasing and concave utility function. 
There is a considerable body of literature exploring different applications of various network utility functions such as network-wide proportional fairness \cite{yigal07fairness}, network-wide max-min fairness \cite{rasek20joint}, network-wide logarithmic utility for balanced load distribution \cite{ye13user}, and $\alpha$-optimal user association \cite{kim12distributed}. 
The particular choice of $U(\cdot)$ is at the liberty of the network designer and design requirements. 
In our simulation we will use the logarithmic utility function, $U(\cdot) = \log(\cdot)$, which maximizes the product of rates and achieves a healthy balance between network sum-rate and fairness, and 
is a common choice for network design and evaluation \cite{srikant2014comnets}. 

\textbf{Optimization problem.}
With our constraints and objective mathematically defined, the network design problem can be formulated as follows, which aims to find the arrival rates $\blam$ and resource allocations $\bmu$ that maximize the sum utility of the network subject to our constraints.
\begin{subequations} \label{opt:max_delay}
\begin{align}
\underset{\blam, \bmu}{\mathrm{maximize}} ~&\sum_{m=1}^{M} U(\lambda_m) \label{opt:max_delayobj}\\
\st~
& \blam \geq\bm{0},~\bm{0}\leq \bmu \leq \bm{1} \label{opt:max_delayvar_intervals}\\
& \bC\bmu - \bF\blam > \bm{0} \label{opt:max_delayrate_feas_cons}\\ 
& \bG\bmu \leq \bm{1} \label{opt:max_delayresource_cons}\\
& \sum_{(u,v)\in \sfR(m)}\log\left(1 - e^{-(c_{u,v}\mu_{u,v} - (\bF\blam)_v)\delta/h_m}\right) \geq \log(\eta), \quad m = 1, 2, \dots, M \label{opt:max_delaydelay_cons}
\end{align}
\end{subequations}
The objective \eqref{opt:max_delayobj} is to maximize the network utility with respect to $\blam$. 
Constraints \eqref{opt:max_delayvar_intervals} are the fundamental properties of $\blam$ and $\bmu$, and constraint \eqref{opt:max_delayrate_feas_cons} ensures that the queues are stable and do not accumulate to infinity. 
Along with the quality of the wireless environment, the capacity matrix $\bC$ depends on whether the \iab nodes are \hd- or \fd-capable. 
Constraint \eqref{opt:max_delayresource_cons} represents the scheduling constraint on each BS, and the scheduling matrix $\bG$ depends on if the \iab nodes are upgraded from \hd to \fd. 
Finally, \eqref{opt:max_delaydelay_cons} represents the probabilistic latency constraint.

\begin{remark}
Having \fd capability at the \iab nodes relaxes the scheduling constraints \eqref{opt:max_delayresource_cons} on the \iab nodes, and they can receive data from their parent while simultaneously transmitting data to one of their children.
This introduces a variety of potential network gains. 
First, a relaxed scheduling constraint provides more scheduling opportunities to the backhaul links. 
This expands the support of the network, also termed its \textit{throughput region}, and
as a consequence, an \fd-\iab network can support higher arrival rates per UE, denoted by the elements of $\blam$, while still satisfying \eqref{opt:max_delaydelay_cons}. 
Second, an \iab network with \fd \iab nodes can support tighter delay constraints. 
In other words, relaxing the constraint \eqref{opt:max_delayresource_cons} by upgrading to \fd \iab nodes, an \iab network can achieve latency targets $\delta$ that may have been infeasible for the equivalent \hd-\iab network.
Moreover, since delay can grow arbitrarily as a network's operating point approaches the boundary of its throughput region, a seemingly minor 
expansion in the throughput region can translate to a considerable reduction in latency.
Third, with the expansion of the throughput region and the feasible values of $\delta$, \fd-\iab can meet throughput targets with \textit{deeper} networks and support more hops. 
This is advantageous for operators as they can then provide comparable quality-of-service with fewer fiber-connected donor nodes, reducing infrastructure costs. 
\end{remark} 
\section{Minimum Feasible Delay Threshold and Latency Gain}
\label{Sec:PerAnalysis} 

The optimization problem in \eqref{opt:max_delay} provides a framework to study these network trade-offs and meet system requirements. 
We now seek analytical expressions and insights using the optimization framework in \secref{Sec:ProbForm}. 
Specifically, we aim to answer the following questions. \textit{(i) What is the minimum delay threshold $\delta^*$ that an \iab network parameterized by $\bF$, $\bG$, $\bC$ can support? (ii) What is the gain in $\delta^*$ for \fd-\iab over \hd-\iab? (iii) What is the maximum number of hops feasible for an \iab network and how does it vary with $\delta^*$? (iv) How does $\delta^*$ vary with the minimum per-UE throughput requirement $\minlam$?} 
With these questions in mind, we formulate and solve a linear program to find the minimum delay threshold, subject to the constraints from \eqref{opt:max_delay}. 

\subsection{Minimum Feasible Delay}
Latency is a key performance metric in modern networks.
5G cellular networks, for example, are designed to support end-to-end packet delays on the order of milliseconds for mission critical and tactile internet applications \cite{fettweis2014tactile}. 
In \eqref{opt:max_delay}, the delay threshold of a network is captured by $\delta$ and constraint \eqref{opt:max_delaydelay_cons}. 
For a network designer, it is useful to know the minimum delay a network can support and how it relates network parameters, such as its routing matrix $\bF$, link capacities $\bC$, and scheduling matrix $\bG$.
Mathematically, this is equivalent to finding the minimum $\delta$ such that problem \eqref{opt:max_delay} is feasible.

Minimizing $\delta$ without any constraints on $\blam$, however, could lead to solutions that are practically undesirable where UEs receive zero throughput. 
We address this by introducing a new constraint 
\begin{align}
 \blam \geq \minlam \bm{1}
\end{align}
to ensure a minimum average arrival rate $\minlam$ is met at each UE, which can be tuned by network engineers.
While one can solve for the minimum $\delta$ using numerical solvers, the delay constraint \eqref{opt:max_delaydelay_cons} is non-convex when optimizing over $\delta$, $\blam$, and $\bmu$, and hence intractable. 
This motivates us to substitute \eqref{opt:max_delaydelay_cons} with a tighter constraint by replacing $\bbP[\tilde{D}_m h_m\leq \delta] \geq \eta$ with $\bbP[D_v h_m\leq\delta] \geq \eta, \forall (u,v)\in\sfR(m)$.
Combining all of this leads to the formulation of problem \eqref{opt:delay_feas_lp}, which includes minimizing over $\delta$ and incorporates modified constraints \eqref{opt:delay_feas_lp_var_intervals} and \eqref{opt:delay_feas_lp_delay_cons}.
\begin{subequations}\label{opt:delay_feas_lp}
\begin{align}
\underset{\delta, \bmu, \blam}{\mathrm{minimize}} ~&\delta\\
\st~& \blam \geq \minlam\bm{1},~\bm{0} \leq \bmu \leq \bm{1}, \delta \geq 0 \label{opt:delay_feas_lp_var_intervals}\\
& \bC\bmu - \bF\blam > \bm{0} \label{opt:delay_feas_lp_rate_feas_cons}\\
& \bG\bmu \leq \bm{1} \label{opt:delay_feas_lp_resource_cons}\\
& 1 - e^{-(c_v\mu_v - (\bF\blam)_v)\delta/h_m} \geq \eta \ \forall (u,v)\in\sfR(m), \ \forall m = 1, 2, \dots, M \label{opt:delay_feas_lp_delay_cons}
\end{align}
\end{subequations}
The new latency constraint \eqref{opt:delay_feas_lp_delay_cons} ensures that the per-hop delay is less than $\delta/h_m$, which is a stricter requirement than \eqref{opt:max_delaydelay_cons}, and $h_m$ is as defined in \eqref{eq:max_delay_CDF}.
Rearranging the terms in \eqref{opt:delay_feas_lp_delay_cons} and applying the change of variable $t=-\log(1 - \eta)/\delta$, problem \eqref{opt:delay_feas_lp} can be reformulated as the following linear program.
\begin{subequations} \label{opt:delay_feas_lp_reform}
\begin{align}
\underset{t, \blam, \bmu}{\mathrm{maximize}} ~&t\\
\st~& \blam \geq\minlam\bm{1},~\bm{0} \leq \bmu \leq \bm{1}, ~t\geq 0 \label{opt:delay_feas_lp_reform_var_intervals}\\
& \bG\bmu \leq \bm{1} \label{opt:delay_feas_lp_reform_resource_cons}\\ 
& c_v\mu_v - (\bF\blam)_v \geq t h_m, ~~\forall (u,v)\in\sfR(m), ~ \forall m = 1, 2, \dots, M
\label{opt:delay_feas_lp_reform_delay_cons}
\end{align}
\end{subequations}

\begin{lemma}
\label{lem:min_lam_opt}
The optimal arrival rate vector for \eqref{opt:delay_feas_lp_reform} is $\blam^* = \minlam\bm{1}$.
    \begin{proof}
    Let $(u',v')$ be the bottleneck edge such that $c_{v'}\mu^*_{v'} - (\bF\blam^*)_{v'} = t^*h_{m'}$, for some UE $m'$ and $\lambda^*_{m'}  > \minlam$, where $t^*$ and $\bmu^*$ denote the corresponding optimal points. Since, $c_{v'}\mu^*_{v'} - \minlam(\bF\bm{1})_{v'} > c_{v'}\mu^*_{v'} - (\bF\blam^*)_{v'}$, $\minlam\bm{1}$ relaxes the bottleneck constraint and achieves a higher objective, which contradicts the optimality of $t^*$.
    \end{proof}
\end{lemma}

\begin{remark}
Note that the $\blam^*=\minlam\bm{1}$ is not a unique solution. For the non-bottleneck $(u,v)$ and $m$ such that $c_{v}\mu^*_{v} - \minlam(\bF\bm{1})_{v} > t^*h_{m}$, any arrival rate $\lambda_m$ which satisfies
\begin{align*}
    \minlam < \lambda_m < \frac{c_v\mu^*_{v} - t^*h_m}{(\bF\bm{1})_{v}}
\end{align*}
will not change the optimal $t$, and hence is also optimal.
\end{remark}

\begin{theorem}
\label{thm:feasible_delay_sol_general}
Given the routing matrix $\bF$, scheduling matrix $\bG$, and capacity matrix $\bC$, the optimal solution for \eqref{opt:delay_feas_lp_reform} is given by
\begin{align}
    \label{eq:feasible_delay_sol_general}
    t^* = \underset{k=0,1,\dots,K}{\min}\frac{1 - \minlam\bG_{k,:}\bC^{-1}\bF\bm{1}}{\bG_{k,:}\bC^{-1}\tilde{\bh}},
\end{align}
where $\bG_{k,:}$ is the $k$-th row of $\bG$ and $\tilde{\bh}$ is a $|\cE|\times 1$ vector such that $\tilde{h}_{v} = \underset{m:(u,v)\in\sfR(m)}{\max}h_m$. 
If $t^*$ is less than zero, then \eqref{opt:delay_feas_lp_reform} is infeasible and the \iab network parameterized by $\bF,\bG, \bC$ cannot support a per-UE arrival rate of $\minlam$.

    \begin{proof}
    Using \lemref{lem:min_lam_opt} and substituting $\bmu$ from \eqref{opt:delay_feas_lp_reform_delay_cons} in \eqref{opt:delay_feas_lp_reform_resource_cons}, we get
    \begin{align*}
        \bG_{k,:}\bC^{-1}(t\tilde{\bh} + \minlam\bF\bm{1}) \leq 1, ~~\forall k = 0,1,\dots,K \\
        t \leq \frac{1 - \minlam\bG_{k,:}\bC^{-1}\bF\bm{1}}{\bG_{k,:}\bC^{-1}\tilde{\bh}}, ~\forall k=0,1,\dots,K
    \end{align*}
    Hence, the optimal $t^*$ achieves the tightest inequality and is given by \eqref{eq:feasible_delay_sol_general}.
    \end{proof}
\end{theorem}

\begin{remark}
The minimum feasible delay for the \iab network is given by $\delta^* = -\log(1-\eta)/t^*$.
\end{remark}


\subsection{Latency Gain}
 Given \thmref{thm:feasible_delay_sol_general}, we can compute $\delta^*_{\hdrm}$ and $\delta^*_{\fdrm}$ for \hd and \fd deployments, respectively. 
 The \textit{latency gain} $\ell$ of \fd-\iab over \hd-\iab can then be expressed as
\begin{align}
    \label{eq:latency_gain_general}
    \ell = \frac{\delta^*_{\hdrm}}{\delta^*_{\fdrm}} = \frac{t^*_{\fdrm}}{t^*_{\hdrm}}= \frac{\underset{k=0,1,\cdots,K}{\min}\frac{1 - \minlam(\bG_{\fdrm})_{k,:}\bC_{\fdrm}^{-1}\bF\bm{1}}{(\bG_{\fdrm})_{k,:}\bC_{\fdrm}^{-1}\tilde{\bh}}}{\underset{k=0,1,\cdots,K}{\min}\frac{1 - \minlam(\bG_{\hdrm})_{k,:}\bC_{\hdrm}^{-1}\bF\bm{1}}{(\bG_{\hdrm})_{ k,:}\bC_{\hdrm}^{-1}\tilde{\bh}}}.
\end{align}
Here, $\bG_{\hdrm}$ and $\bG_{\fdrm}$ denote the scheduling matrices for the \hd and \fd deployments, respectively, and $\bC_{\hdrm}$ and $\bC_{\fdrm}$ denote the corresponding the capacity matrices. 
Depending on the quality of \si cancellation and the structure of the \iab routing tree $\cT$, the bottleneck BS---which achieves the minimum in \eqref{eq:feasible_delay_sol_general}---could be different for \fd and \hd deployments. 
Simplified expressions can be obtained if one considers the structure of the routing matrix $\bF$ or particular cases of $\bC$. 

\begin{figure}[t]
    \centering
    \includegraphics[height=0.25\textheight,keepaspectratio]{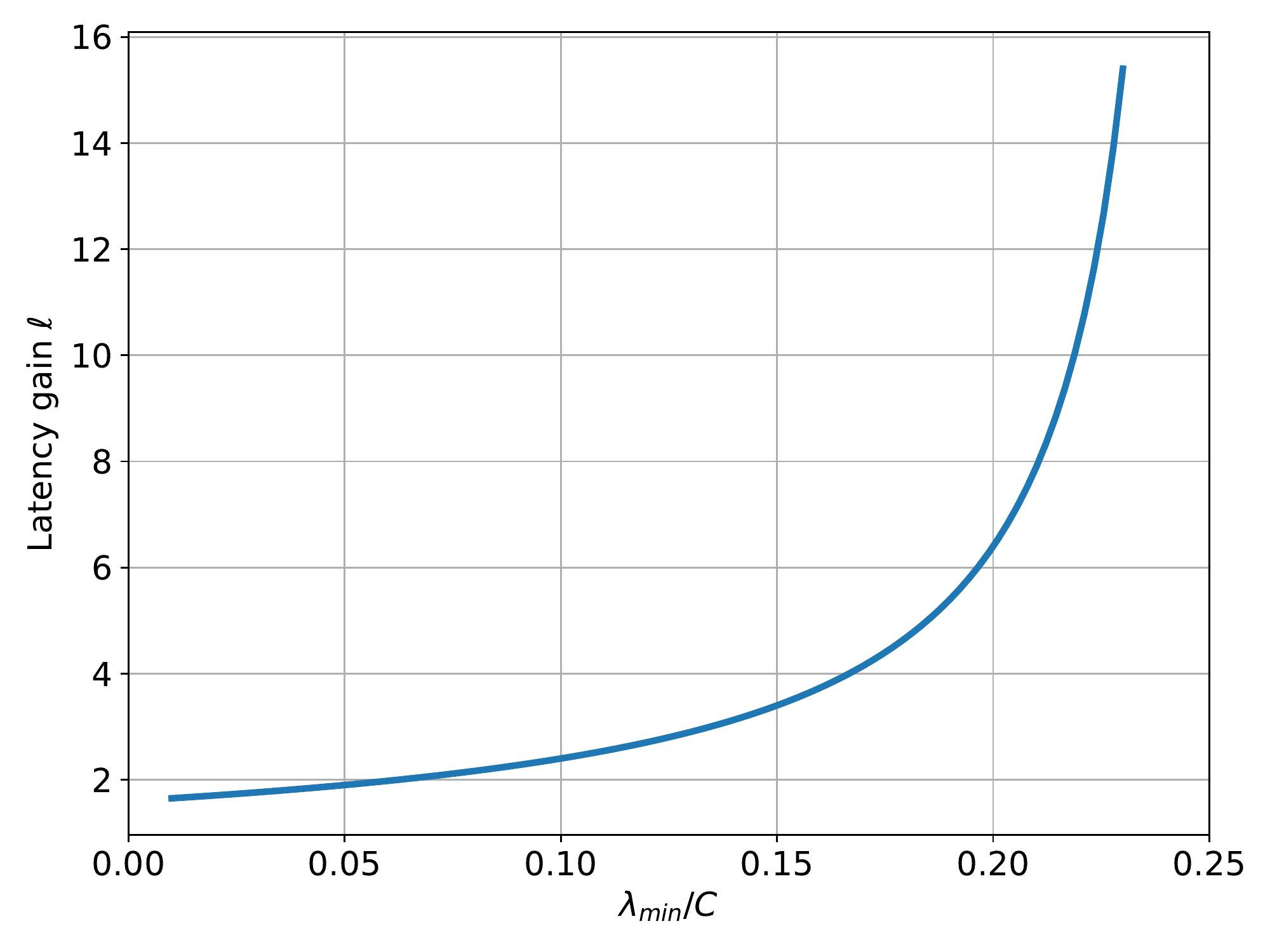}
    \caption{Latency gain for a three-hop line network where all links have capacity $C$ and there exists perfect \si cancellation at \fd \iab nodes. \fd-\iab can support minimum per-user arrival rates infeasible for \hd-\iab and arbitrarily large latency gains can be observed.}
    \label{fig:latency_gain_3node_net}
\end{figure}

To illustrate a simple example, we refer to the \iab routing tree given in \figref{fig:3HopNet} to have
\begin{align}
    \bF = \begin{bmatrix} 
            1 & 0 & 0 & 0 & 0\\
            0 & 1 & 1 & 0 & 0\\
            0 & 1 & 0 & 1 & 1
          \end{bmatrix}^T
\end{align}
and $\tilde{\bh} = [1, 3, 2, 3,3]^T$ (defined in \eqref{eq:feasible_delay_sol_general}).
The \hd and \fd scheduling matrices are as given in \eqref{eq:scheduling_mat_ex}, respectively. 
If we assume for this example that all edges have equal capacity $C$ and $\bC_{\hdrm} = \bC_{\fdrm} = C\bI$, then using \thmref{thm:feasible_delay_sol_general} we have
\begin{align}
    t^*_{\hdrm} &= \frac{C - 4\minlam}{8}\\
    t^*_{\fdrm} &= \begin{dcases}
                        \frac{C - 2\minlam}{5}, &\minlam \leq C/7\\
                        \frac{C - 3\minlam}{4}, & \textrm{else}
                    \end{dcases}
\end{align}
and the latency gain is given by $\ell = t^*_{\fdrm}/t^*_{\hdrm}$,
which is plotted in \figref{fig:latency_gain_3node_net}. 
This simple but insightful example illustrates that \iab with \fd nodes can support $\ell = 2.5$ times tighter latency constraints when the minimum arrival rate to each UE $\minlam$ is $10\%$ of the link capacity $C$.
When $\minlam$ is just over a fifth of $C$, the latency gain is $\ell = 12$ (nearly quintuples) and increases substantially for small increases in $\minlam$ thereafter.
This behavior is explained by the simple fact that \fd-\iab can support higher values of $\minlam$ which are infeasible for \hd-\iab, driving the latency gain $\ell$ toward infinity.

\section{Analytical Results: Latency and Rate of a Line Network}
In the interest of providing simple, intuitive, and tractable expressions that may serve as guidelines for network design, we will present a detailed analysis of a simple but practically relevant \iab deployment: the \textit{line network}.
A routing tree $\cT$ represents a line network if no two BSs in $\cT$ share a parent (e.g., the \iab network shown in \figref{fig:3HopNet}).
We also assume that $\bC_{\hdrm} = \bC_{\fdrm} = \bC$, meaning \fd \iab nodes are capable of perfect self-interference cancellation. 
In \secref{sec:results}, we will show that there exists a self-interference cancellation threshold which, if achieved, is effectively as good as perfect cancellation from a network perspective. 
We assume that all backhaul edges have capacity $R_b$ and all access edges have capacity $R_a$.
We also assume that $R_b > R_a$ because BSs typically have more antennas and benefit from a higher array gain compared to UEs, along with the simple fact that \iab networks are typically designed in this fashion this since they are fundamentally backhaul-limited.

\subsection{Latency Gain}
In \eqref{eq:latency_gain_general}, we derived latency gain $\ell$ for a \textit{general} \iab routing tree. In the following, we will derive a closed-form expression for \eqref{eq:feasible_delay_sol_general} under both \hd and \fd deployments of a line network, along with the corresponding latency gain $\ell$.
\begin{theorem}
\label{thm:feasible_delay_sol_ra_rb}
Let $\cT$ represent a line network with $K$ \iab nodes and one donor such that each BS supports $w$ UEs, where all backhaul edges have capacity $R_b$ and all access edges have capacity $R_a$. Then, for a feasible $\minlam$ and under perfect \si cancellation, we have
\begin{align}
    t^*_{\hdrm}(\minlam, K) &= 
     \begin{dcases}
        \frac{1 - w\minlam\left(\frac{3}{R_b} + \frac{1}{R_a}\right)}{\frac{2(K+1)}{R_b} + \frac{Kw}{R_a}}, & \minlam \leq \frac{R_a}{4(K+1)\left(\frac{R_a}{R_b}\right)^2 + (2K+3)w\left(\frac{R_a}{R_b}\right) + w} \\
        \frac{1 - w\minlam\left(\frac{2K-1}{R_b} + \frac{1}{R_a}\right)}{\frac{2(K+1)}{R_b} + \frac{2w}{R_a}}, & \mathrm{else}
    \end{dcases} \\
    t^*_{\fdrm}(\minlam, K) &= 
     \begin{dcases}
        \frac{1 - w\minlam\left(\frac{1}{R_b} + \frac{1}{R_a}\right)}{\frac{K+1}{R_b} + \frac{Kw}{R_a}}, & \minlam \leq \frac{R_a}{(K+1)\left(\frac{R_a}{R_b}\right)^2 + (K+1)w\left(\frac{R_a}{R_b}\right) + w} \\
        \frac{1 - w\minlam\left(\frac{K}{R_b} + \frac{1}{R_a}\right)}{\frac{K+1}{R_b} + \frac{w}{R_a}}, & \mathrm{else}
    \end{dcases}.
\end{align}

\begin{proof}
See \appref{app:feasible_delay_sol_ra_rb}.
\end{proof}
\end{theorem}

Note that, since $\cT$ represents a line network, $K+1$ is also the number of hops between the donor and the furthest UE. 
It is interesting to note that the bottleneck BS is either the first or the last \iab node. 
This can be attributed to the function $f(\cdot)$ (see \appref{app:feasible_delay_sol_ra_rb}) and its property that it is either non-increasing or non-decreasing depending on whether $\minlam$ is above or below a threshold value given in \thmref{thm:feasible_delay_sol_ra_rb}. 
On close inspection, we see that $f$ is the ratio of the fraction of time the BS is idle and the time it takes for the BS to serve one unit of the total arriving traffic (i.e., $\minlam\bF\bm{1}$) under the delay constraints. 
The idle time is an increasing function of $k$ and a decreasing function of $\minlam$, whereas $\minlam\bF\bm{1}$ is an increasing function of $k$. 
This explains the peculiar property of $f$. 
One can also interpret $f$ as the product of the service rate of the BS and the fraction of time it is idle---in some sense it is the effective service rate. 
The inverse of which is the average delay experienced by the arriving flow at the BS. 
The objective of \eqref{opt:delay_feas_lp_reform} (or equivalently \eqref{opt:delay_feas_lp}) is to find the \textit{bottleneck} BS with the maximum service time within the resource allocation constraints. 
The delay threshold $\delta$ an \iab network can support is limited by this bottleneck BS.
\begin{corollary}
\label{cor:latency_gain_ra_rb}
For the network in \thmref{thm:feasible_delay_sol_ra_rb}, the latency gain is given by 
\begin{align}
    \ell = 
    \begin{dcases}
        \frac{1 - w\minlam\left(\frac{1}{R_b} + \frac{1}{R_a}\right)}{1 - w\minlam\left(\frac{3}{R_b} + \frac{1}{R_a}\right)}
        \frac{\frac{2(K+1)}{R_b} + \frac{Kw}{R_a}}{\frac{K+1}{R_b} + \frac{Kw}{R_a}}, ~ &\minlam \leq \Lambda_1 \\
        \frac{1 - w\minlam\left(\frac{1}{R_b} + \frac{1}{R_a}\right)}{1 - w\minlam\left(\frac{2K-1}{R_b} + \frac{1}{R_a}\right)}
        \frac{\frac{2(K+1)}{R_b} + \frac{2w}{R_a}}{\frac{K+1}{R_b} + \frac{Kw}{R_a}}, ~ &\Lambda_1 < \minlam \leq \Lambda_2 \\
        2\frac{1 - w\minlam\left(\frac{K}{R_b} + \frac{1}{R_a}\right)}{1 - w\minlam\left(\frac{2K-1}{R_b} + \frac{1}{R_a}\right)}, ~ &\mathrm{else}
    \end{dcases}
\end{align}
where
\begin{align*}
    \Lambda_1 &= \frac{R_a}{4(K+1)\left(\frac{R_a}{R_b}\right)^2 + (2K+3)w\left(\frac{R_a}{R_b}\right) + w} \quad
    \Lambda_2 &= \frac{R_a}{(K+1)\left(\frac{R_a}{R_b}\right)^2 + (K+1)w\left(\frac{R_a}{R_b}\right) + w}.
\end{align*}
\begin{proof}
This is found directly by substituting $t^*$ from \thmref{thm:feasible_delay_sol_ra_rb} into $\delta^* = -\log(1 - \eta) / t^*$.
\end{proof}
\end{corollary}

\subsection{Maximum Network Depth}
In this section, we present analytical expressions for the maximum depth $K^{\maxrm}$ of a line network that can support given latency and throughput targets.
Note that the result of \thmref{thm:feasible_delay_sol_ra_rb} can also be written as 
\begin{align}
    \label{eq:t_opt_hd_K_breaks}
    t^*_{\hdrm}(\minlam, K) &= 
     \begin{dcases}
        \frac{1 - w\minlam\left(\frac{3}{R_b} + \frac{1}{R_a}\right)}{\frac{2(K+1)}{R_b} + \frac{Kw}{R_a}}, & K \leq \frac{\frac{R_a}{\minlam} - w - \frac{R_a}{R_b}\left(4\frac{R_a}{R_b} + 3w\right)}{2\frac{R_a}{R_b}\left(2\frac{R_a}{R_b} + w\right)} \\
        \frac{1 - w\minlam\left(\frac{2K-1}{R_b} + \frac{1}{R_a}\right)}{\frac{2(K+1)}{R_b} + \frac{2w}{R_a}}, & \mathrm{else}
    \end{dcases} \\
    \label{eq:t_opt_fd_K_breaks}
    t^*_{\fdrm}(\minlam, K) &= 
     \begin{dcases}
        \frac{1 - w\minlam\left(\frac{1}{R_b} + \frac{1}{R_a}\right)}{\frac{K+1}{R_b} + \frac{Kw}{R_a}}, & K \leq \frac{\frac{R_a}{\minlam} - w}{\frac{R_a}{R_b}\left(\frac{R_a}{R_b} + w\right)} - 1\\
        \frac{1 - w\minlam\left(\frac{K}{R_b} + \frac{1}{R_a}\right)}{\frac{K+1}{R_b} + \frac{w}{R_a}}, & \mathrm{else}.
    \end{dcases}
\end{align}

Let $K^{\maxrm}_{\hdrm}$ and $K^{\maxrm}_{\fdrm}$ represent the maximum number of \iab nodes for the \hd and \fd line networks, respectively, that can guarantee a target delay threshold $\delta_{\mathrm{target}}$ and minimum arrival rate $\minlam$. 
For notational convenience, let $\kappa_{\hdrm}$ and $\kappa_{\fdrm}$ denote the break points from \eqref{eq:t_opt_hd_K_breaks} and \eqref{eq:t_opt_fd_K_breaks} as follows.
\begin{align}
    \kappa_{\hdrm} = \frac{\frac{R_a}{\minlam} - w - \frac{R_a}{R_b}\left(4\frac{R_a}{R_b} + 3w\right)}{2\frac{R_a}{R_b}\left(2\frac{R_a}{R_b} + w\right)} \quad
    \kappa_{\fdrm} = \frac{\frac{R_a}{\minlam} - w}{\frac{R_a}{R_b}\left(\frac{R_a}{R_b} + w\right)} - 1
\end{align}

\begin{theorem}
\label{thm:k_max}
Let $\cT$ represent a line network where all backhaul edges have capacity $R_b$, all access edges have capacity $R_a$, and each BS supports $w$ UEs. 
Suppose $K^{\maxrm}_{\hdrm}+1$ and $K^{\maxrm}_{\fdrm}+1$ are the maximum number of hops that can still meet a target delay threshold $\targetdelta$ and a minimum arrival rate $\minlam$ for \fd-\iab and \hd-\iab deployments, respectively. 
Then, 
\begin{align}
    \label{eq:k_max_hd}
    K^{\maxrm}_{\hdrm} &= 
    \begin{dcases}
        \floor{\frac{1 - w\minlam\left(\frac{3}{R_b} + \frac{1}{R_a}\right) - \frac{2\ttar}{R_b}}{2\ttar\left(\frac{2}{R_b} + \frac{w}{R_a}\right)}}, & \ttar \leq t^*_{\hdrm}(\minlam,\kappa_{\hdrm})\\
        \floor{\frac{1 - w\minlam\left(\frac{1}{R_a} - \frac{1}{R_b}\right) - 2\ttar\left(\frac{1}{R_b} + \frac{w}{R_a}\right)}{\frac{2\ttar}{R_b} + \frac{2w\minlam}{R_b}}}, & \mathrm{else}
    \end{dcases} \\
    \label{eq:k_max_fd}
    K^{\maxrm}_{\fdrm} &=
    \begin{dcases}
        \floor{\frac{1 - w\minlam\left(\frac{1}{R_b} + \frac{1}{R_a}\right) - \frac{\ttar}{R_b}}{\ttar\left(\frac{1}{R_b} + \frac{w}{R_a}\right)}}, & \ttar \leq t^*_{\fdrm}(\minlam,\kappa_{\fdrm}) \\
        \floor{\frac{1 - w\minlam\frac{1}{R_a} - \ttar\left(\frac{1}{R_b} + \frac{w}{R_a}\right)}{\frac{\ttar}{R_b} + \frac{w\minlam}{R_b}}}, & \mathrm{else}
    \end{dcases}
\end{align}
where $\ttar = {-\log(1-\eta)} / {\targetdelta}$.

\begin{proof}
Note that $t^*_{\hdrm}$ and $t^*_{\fdrm}$ are continuous and non-increasing in $\minlam$ and $K$. 
If $\delta^* \leq \targetdelta$ then $t^* \geq \ttar$. 
Choosing the appropriate branch of the piece-wise defined $t^*$ in \eqref{eq:t_opt_hd_K_breaks} and \eqref{eq:t_opt_fd_K_breaks}, and solving for $K$ by setting $t^* \geq \ttar$, we get the above results.
\end{proof}
\end{theorem}

\begin{figure}[t!]
    \centering
    \subfloat[As a function of $\minlam$ for $\targetdelta=10$ msec/packet.]{%
        \includegraphics[width=3in,keepaspectratio]{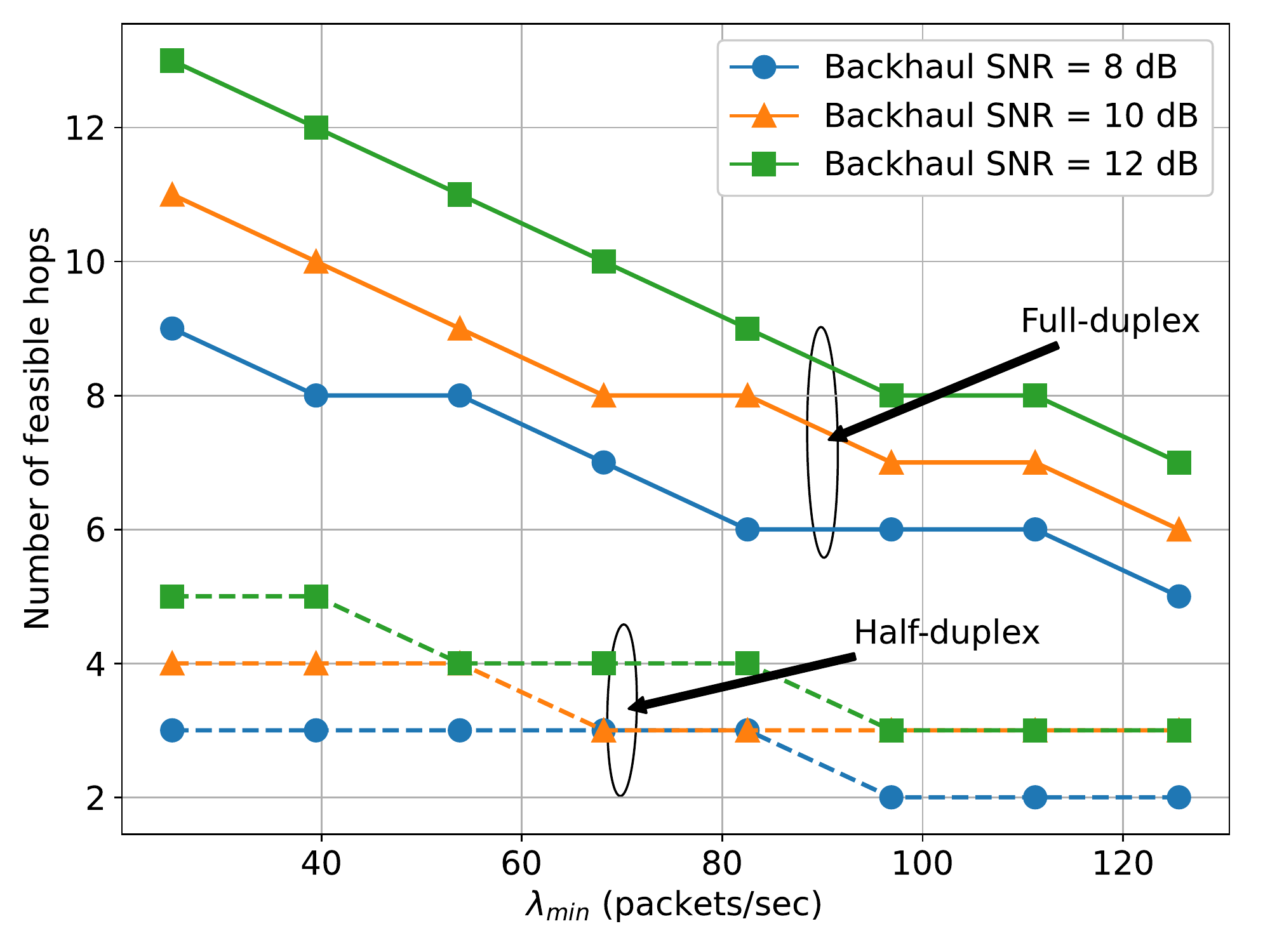}%
        \label{fig:kmax_vs_minlam}}
    \quad
    \subfloat[As a function of $\targetdelta$ for $\minlam=125$ packets/sec.]{%
        \includegraphics[width=3in,keepaspectratio]{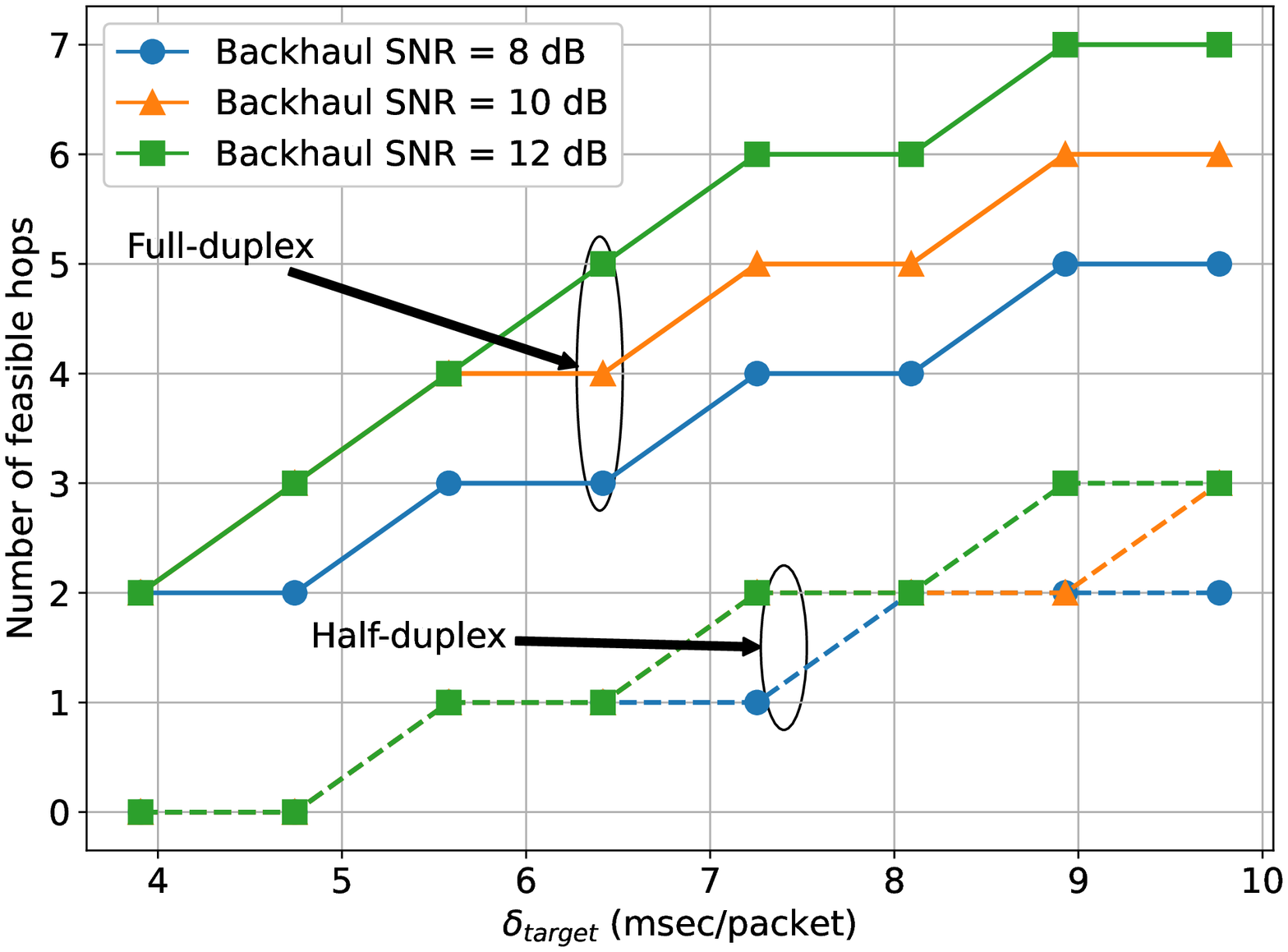}%
        \label{fig:kmax_vs_delta}}
    \caption{The number of feasible hops $K^{\maxrm}$ for \fd-\iab and \hd-\iab as a function of (a) minimum service rate $\minlam$ and (b) $\targetdelta$ for various backhaul \glspl{snr}. \fd-\iab can support more hops while meeting the throughput and latency constraints and the marginal return of increasing backhaul \snr is more compared to \hd-\iab.}
    \label{fig:kmax}
\end{figure}

In \figref{fig:kmax_vs_minlam} we present the effect of $\minlam$ on maximum feasible network depth $K^{\maxrm} + 1$ as computed in \thmref{thm:k_max}, for different backhaul \snr and access \snr of $5$ dB.
\thmref{thm:k_max} assumes perfect \si cancellation and hence $R_b$ and $R_a$ are computed by applying the Shannon formula to \snr. 
We evaluate \eqref{eq:k_max_fd} and \eqref{eq:k_max_hd} for a packet size of $10$ KB, bandwidth $W=100$ MHz, and $w=5$ UEs per BS.
Solid lines denote $K^{\maxrm}_{\fdrm}+1$ and dashed lines denote $K^{\maxrm}_{\hdrm}+1$. 
The plot shows that \fd-\iab can increase maximum feasible network depth by about three times for smaller throughput targets, compared to its \hd counterpart. 
The plot also shows that a $2$ dB increment in the backhaul \snr typically translates to an increment of about two hops in $K^{\maxrm}_{\fdrm}$. 
The same increment in backhaul \snr typically results in an increment of one hop in $K^{\maxrm}_{\hdrm}$, and sometimes no increment at all. In other words, the marginal return of improving the backhaul link quality is more for \fd-\iab. \figref{fig:kmax_vs_delta} shows the variation of $K^{\maxrm}$ with the latency target $\targetdelta$ for fixed $\minlam$ and different backhaul \snr. The plot shows similar trends as \figref{fig:kmax_vs_minlam} and the marginal return of increasing backhaul \snr is more profound.
\thmref{thm:k_max} can also be used to derive a closed-form expression for \textit{hop gain} by following steps similar to \corref{cor:latency_gain_ra_rb}. 



\section{Simulation Results}
\label{sec:results}

In addition to the analytical results presented in the previous section, we also evaluate \fd-\iab network performance---in both line and more general multihop networks---through Monte Carlo simulation and compare it against its \hd counterpart.

\subsection{Channel Model and Signal Propagation}
\label{subsec:channel_model}
We model \mmw channels using the Saleh-Valenzuela-based model \cite{heath16overview}. 
Let $N_{\txrm}$ and $N_{\rxrm}$ denote the number of antennas at the transmitter and receiver respectively. 
The $N_{\rxrm} \times N_{\txrm}$ channel matrix $\bH$ can be written as 
\begin{align}
    \label{eq:channel_matrix}
    \bH &= \sqrt{\frac{1}{\nrays \nclusters}}\sum_{i=1}^{\nclusters}\sum_{j=1}^{\nrays} h_{i,j} \ba_{\rxrm}(\aoa_{i,j})\ba_{\txrm}(\aod_{i,j})^\dagger
\end{align}
which is simply the composition of propagating $\nclusters$ clusters, each having $\nrays$ rays. 
We denote by $h_{i,j} \sim \cC\cN(0,1)$ the complex gain of the $j$-th ray in the $i$-th cluster. 
$\aoa_{i,j}$ is the angle of arrival of the ray at the receiver, and $\aod_{i,j}$ is the angle of departure at the transmitter. 
The scalar in front of the summation handles power normalization. The vectors $\ba_{\txrm}(\cdot)$ and $\ba_{\rxrm}(\cdot)$ represent the array response vectors at the transmitter and the receiver, respectively. 


We assume that the network is noise-limited and interference from neighboring BSs is negligible. 
Hence, received signals are corrupted by additive noise and---at \fd \iab nodes---by residual \si remaining after cancellation. 
Due to lack of measurements and characterization of the \mmw \si channel, we do not assume a particular model for the self-interference channel and our contribution does not rely on specific characteristics of a model. 
Instead, we evaluate our system based directly on \rinr, which captures the degree of \textit{residual} \si plaguing a desired receive signal after analog, digital, and/or spatial cancellation. 
The \sinr at a \fd \iab node $v$ in such a case is expressed as
\begin{align}
    \msinr_v &= \frac{\msnr_{v}}{\mrinr_{v} + 1}
\end{align}
where $\mrinr_v$ is the \rinr at the receiver due to \si and $\msnr_v$ is the \snr the equivalent \hd \iab node observes, written as
\begin{align}
    \msnr_{v} &= \frac{P_{\txrm}L_{v} \bars{\bw_{\rxrm}^\dagger \bH_{v} \bff_{\txrm}}^2}{\sigma^2\cdot W}.
\end{align}
Here, $P_{\txrm}$ denotes the transmit power, $L_{v}$ is the inverse pathloss, $\bH_{v}$ is the channel matrix, $\bff_{\txrm}$, and $\bw_{\rxrm}$ are the transmit and receive beamforming vectors, $W$ is the system bandwidth, $\sigma^2$ is the noise power spectral density. 
The \snr at each UE can be expressed analogously.

\subsection{Network Topology and Channel Parameters}
\label{subsec:eval_setting}
\begin{figure}[t]
    \centering
    \includegraphics[height=1.75in,
                    keepaspectratio]{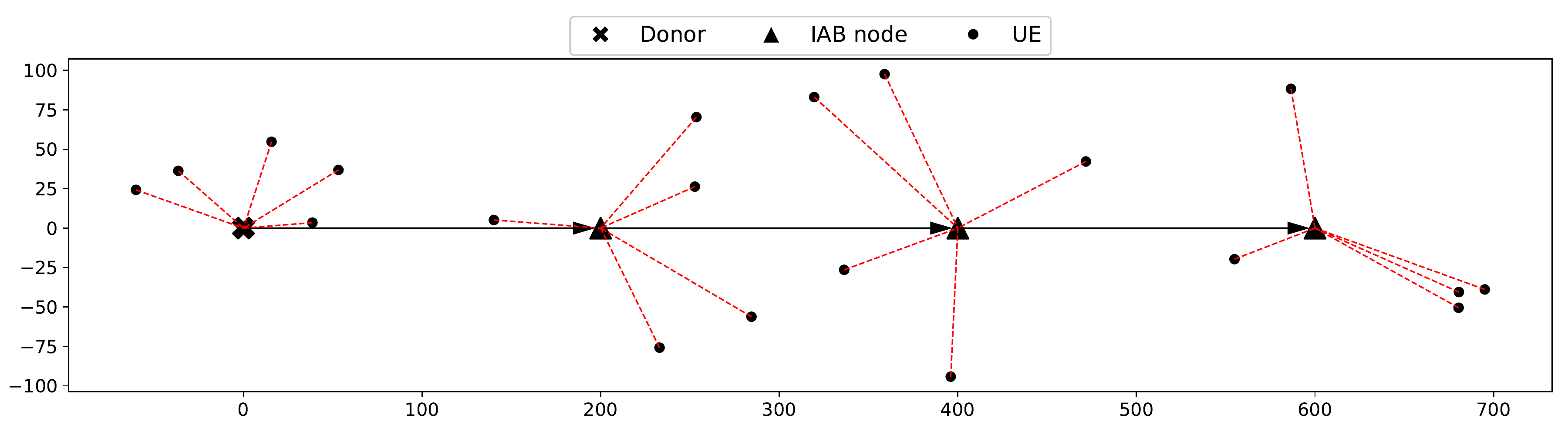}
    \caption{A realization of the simulated line network with depth $K + 1 = 4$, having one donor and three \iab nodes. Five UEs are randomly dropped around the donor and each \iab node.}
    \label{fig:4node_routing_tree}
\end{figure}
Using the proposed optimization framework, we evaluate the gain of \fd-\iab over \hd-\iab for various depths and topologies.
We begin with line networks with depths $K+1=2$, $K+1=3$, and $K+1=4$, and then evaluate a more general \textit{two-child tree} network where the donor and each first-hop \iab node supports two child BSs.
The \iab nodes are separated by $200$ m, the BSs transmit at $30$ dBm, and each BS is equipped with a \ula with $\nbs = 64$ antennas. 
Each UE has a $16$-element \ula. 
Each BS serves five UEs which are dropped uniformly in a $100$ m radius around the BS.
A sample drop for one donor and three \iab nodes---a network depth of $K+1=4$---line network is shown in \figref{fig:4node_routing_tree} and the two-child tree network is illustrated in \figref{fig:symmetric_tree}.
The channels between pairs of devices are generated using \eqref{eq:channel_matrix}, and $\bff$ and $\bw$ are drawn from a \dft codebook to maximize \snr (i.e., codebook-based beam alignment). 
For each Monte Carlo iteration, we drop a new set of UEs and generate channels such that $\nrays$ is drawn uniformly from the interval $[1, 10]$ and $\nclusters$ is drawn from the interval $[1, 6]$ \cite{mfm_arxiv}. 
The system operates at a carrier frequency of $30$ GHz with $100$ MHz bandwidth. 
For the pathloss, we use the model adopted in the 3GPP standard \cite[page 26]{3gpp_tr_38901} for urban environment (UMa), which incorporates the effects of blockage, the multi-slope nature of the pathloss, and the elevation difference between the BSs and the users, through parameters that have been fit to real-world data. 
The additive noise power spectral density is $-174$ dBm/Hz plus a $10$ dB noise figure.

Packets of size $10$ KB arrive at the donor according to a Poisson process like the file transfer protocol (FTP) model 3 \cite{3gpp_tr36889}. 
We divide the capacity of each edge by the mean packet size and normalize it to packets/sec, leading to units of delay in sec/packet. 
Note that these results can be transformed to any packet size by appropriate scaling.
We use CVXPY \cite{cvxpy1} to solve our network design for $\eta=0.9$ (i.e., $90\%$ of the packets must be delivered to the destination UE within the target delay threshold $\delta$).

\subsection{Rate and Latency Gain in Simulated Line Networks}
\label{subsec:rate_latency_gain}
Using \eqref{opt:max_delay}, we will now discuss the effect of deploying \fd-\iab on network throughput, under the logarithmic utility function $U(\cdot) = \log(\cdot)$.
As discussed earlier, the logarithmic utility achieves a healthy balance between network sum-rate and fairness, unlike pure sum-rate maximization which would result in greedy solutions that are unfair and practically undesirable, such as allocating all the resources to a first-hop UE with the best channel.

\begin{figure}
    \centering
    \subfloat[At last-hop for various network depths.]{%
        \includegraphics[width=3in,keepaspectratio]{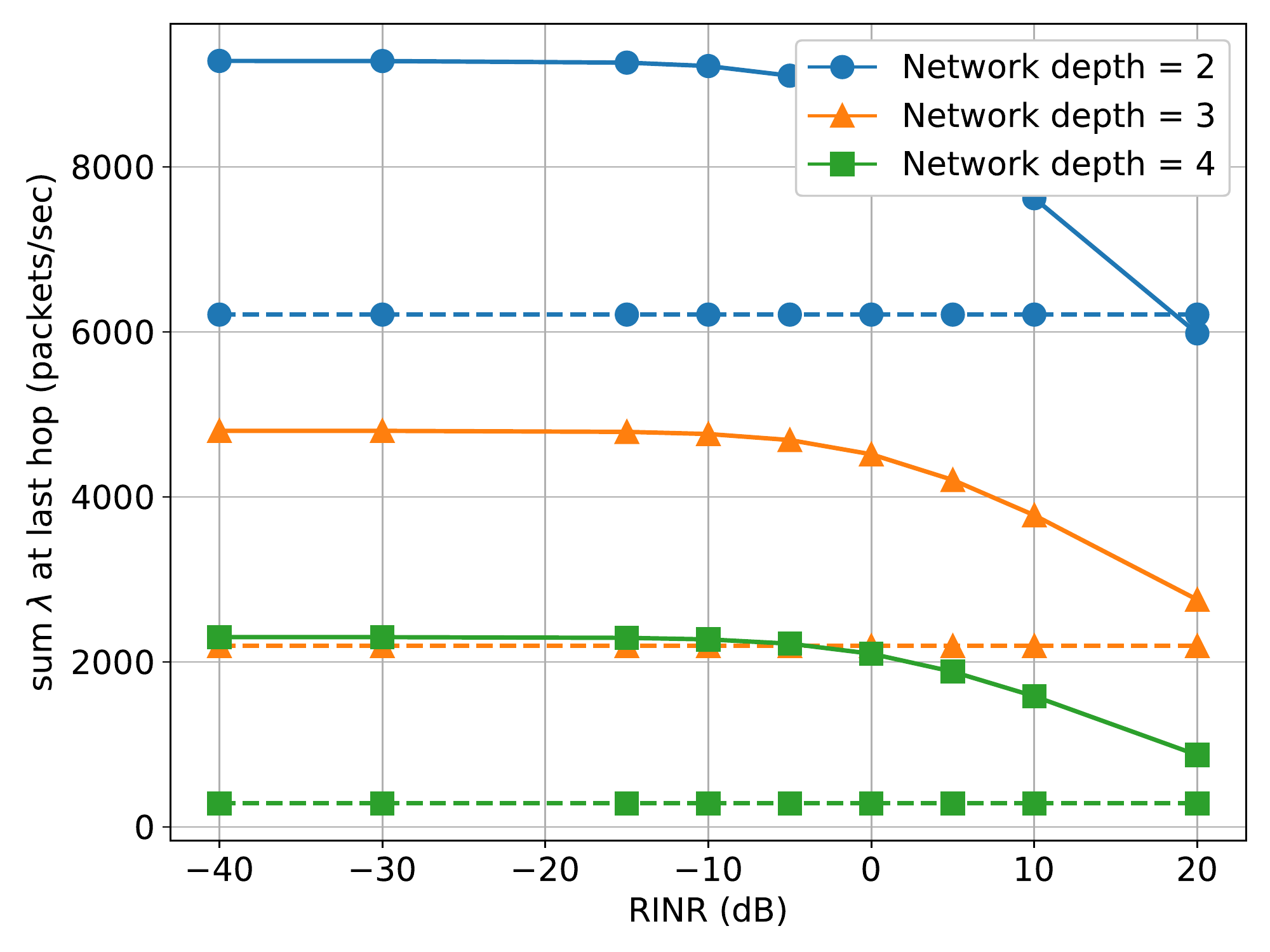}%
        \label{fig:lambda_net_depth}}
    \quad
    \subfloat[At various hops for network depth $K+1=4$.]{%
        \includegraphics[width=3in,keepaspectratio]{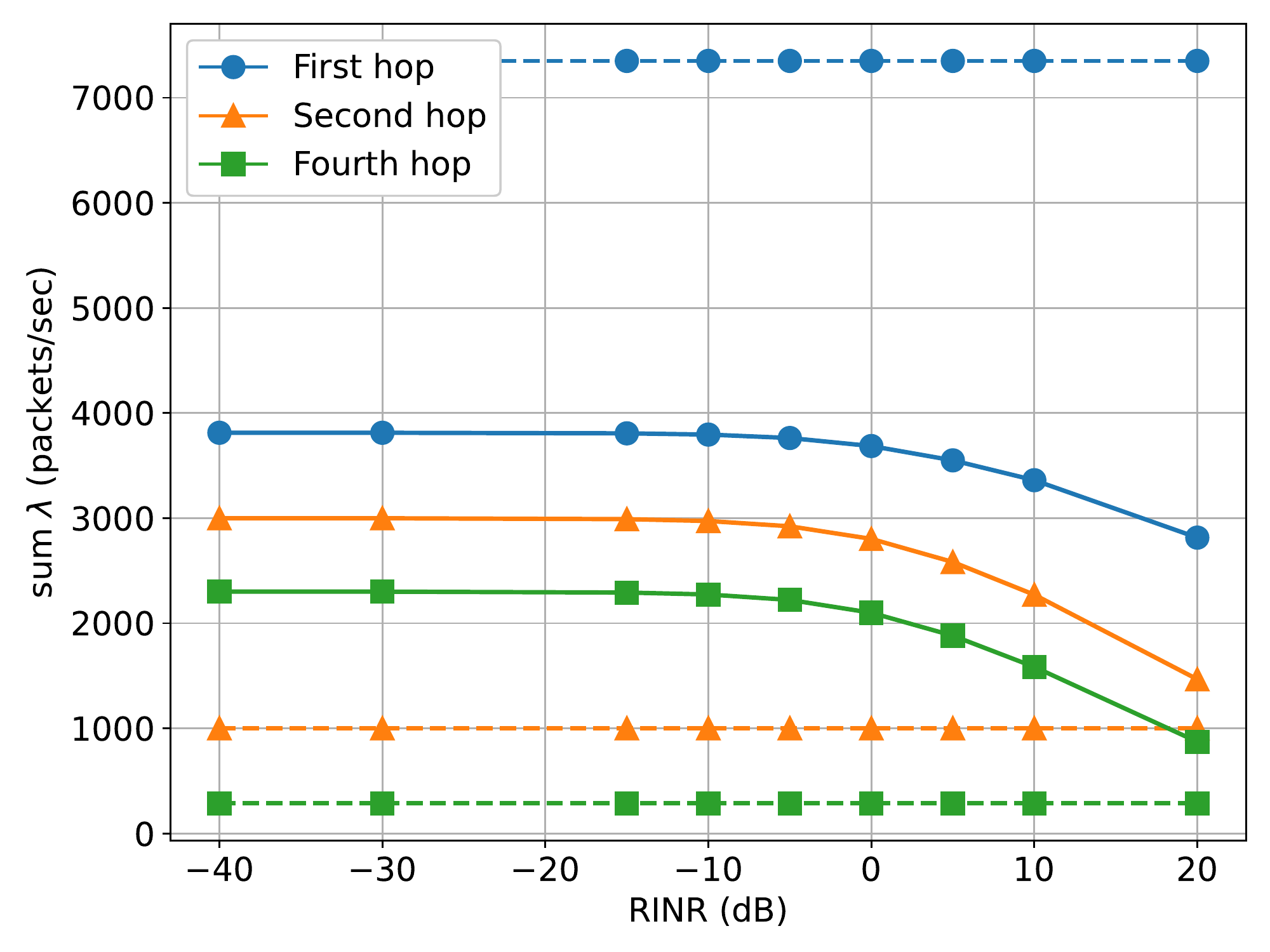}%
        \label{fig:lambda_per_hop}}
    \caption{Sum-rate as a function of \rinr (a) at last-hop for various network depths and (b) at various hops in a network having depth $K+1=4$, where the delay threshold is $\delta = 3.5$ msec/packet. Solid lines represent \fd-\iab and dashed lines represent \hd-\iab. Supporting practical data rates for users at the third and fourth hops is feasible with \fd-\iab.}
    \label{fig:lambda_hd_vs_fd}
\end{figure}

\begin{figure}
    \centering
    \subfloat[At last-hop for various network depths.]{%
        \includegraphics[width=3in,keepaspectratio]{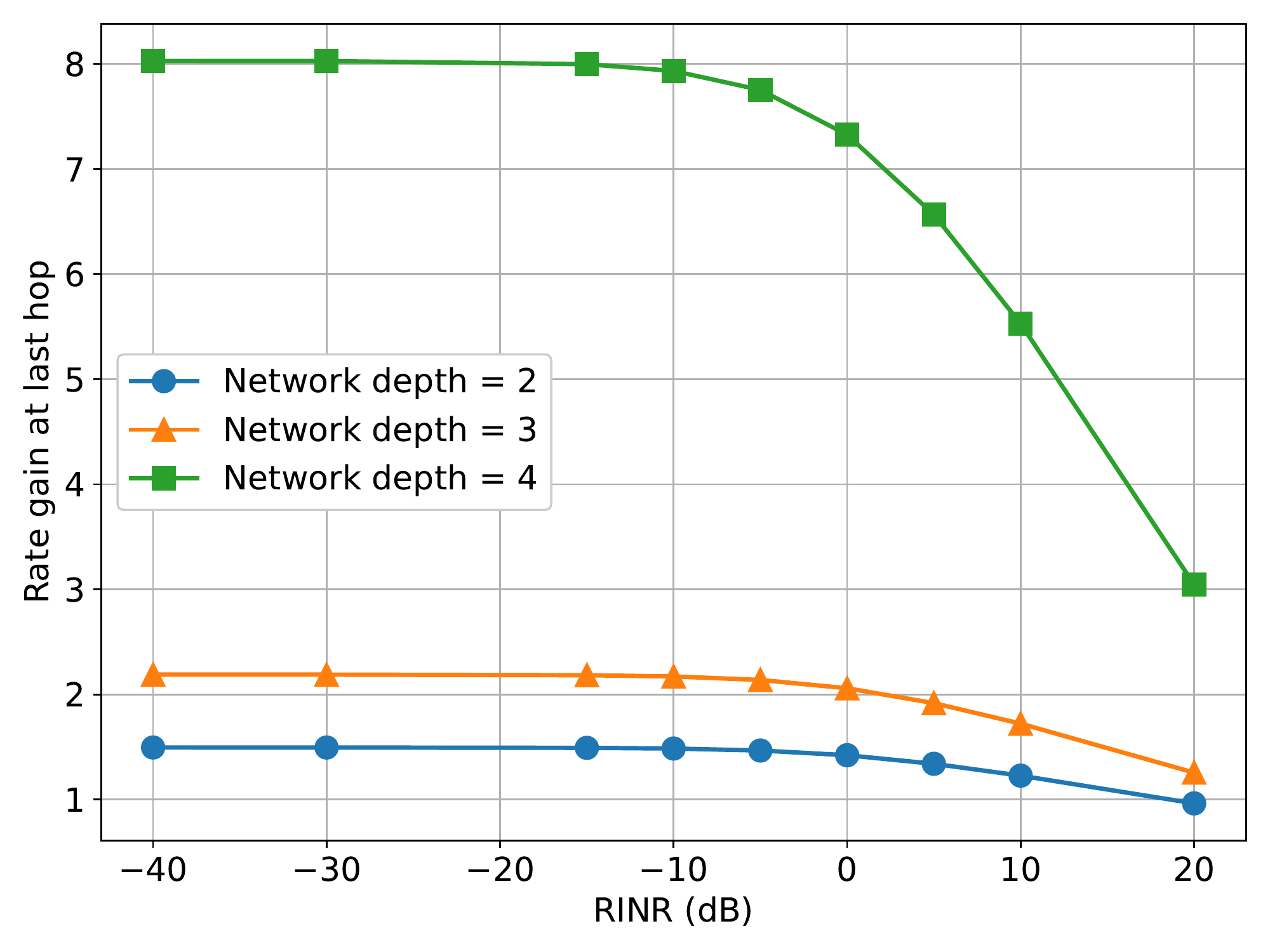}%
        \label{fig:rate_gain_net_depth}}
    \quad
    \subfloat[At various hops for network depth $K+1=4$.]{%
        \includegraphics[width=3in,keepaspectratio]{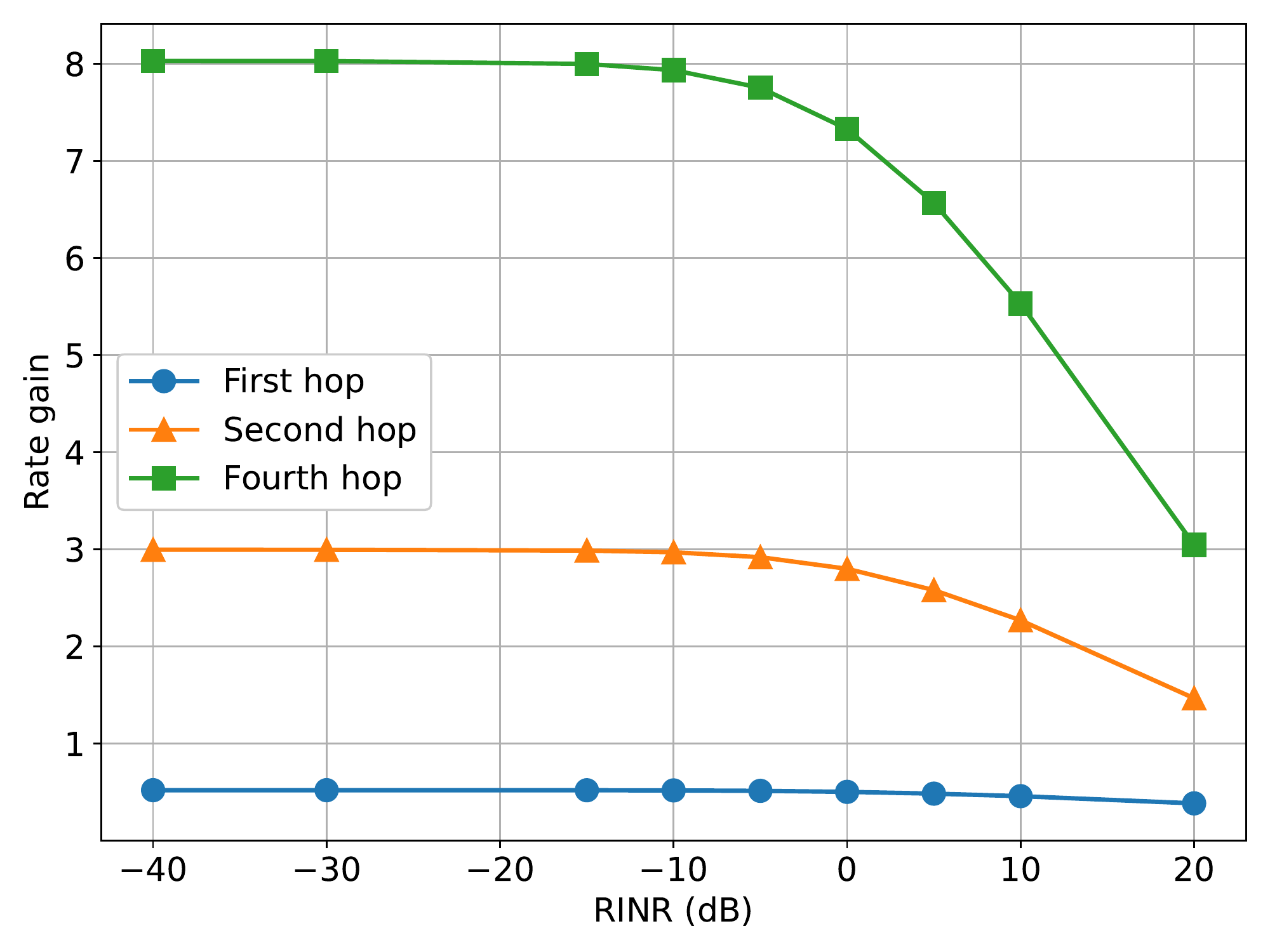}%
        \label{fig:rate_gain_per_hop}}
    \caption{From \figref{fig:lambda_hd_vs_fd}, the rate gain as a function of \rinr (a) at last-hop for various network depths and (b) at various hops in a network having depth $K+1=4$, where the delay threshold is $\delta = 3.5$ msec/packet. As \rinr is decreased, the network observes diminishing gains before saturating. Appreciable gains can be seen for \rinr well above $0$ dB, where self-interference is as strong as noise.}
    \label{fig:rate_gain_hd_vs_fd}
\end{figure}


\textbf{Rate gain with imperfect \si cancellation.} 
For various depths of the described line network, we solve \eqref{opt:max_delay} for both \hd and \fd modes to obtain the optimal arrival rate vectors $\blam_{\hdrm}$ and $\blam_{\fdrm}$. 
Comparing \fd to \hd, the rate gain for UE $m$ is defined as $(\blam_{\fdrm})_m / (\blam_{\hdrm})_m$. 
Let $\cM_i$ denote the set of UEs $i$ hops from the donor. 
Then, the sum-rate at $i$-th hop is defined as $\sum_{\cM_i}\blam_m$ and the rate gain is defined as $\frac{\sum_{\cM_i}(\blam_{\fdrm})_m}{\sum_{\cM_i}(\blam_{\hdrm})_m}$. 
\figref{fig:lambda_net_depth} shows the last-hop sum-rate and \figref{fig:rate_gain_net_depth} shows the rate gain at the last hop versus \rinr for various network depths. 
Naturally, as \rinr increases, sum-rate and consequently rate gain suffer due to increased residual \si plaguing \fd operation, diminishing the resource gains of \fd over \hd.
Even with self-interference that is ten times stronger than noise (i.e., $\mrinr = 10$ dB), however, appreciable gains over \hd are visible, especially for deeper networks.
Reducing \rinr below $-5$ dB does not yield meaningful rate improvements, as sum-rate saturates, which can drive physical layer \fd design decisions from the perspective of a network operator.


\textbf{Deeper networks benefit more from \fd deployment.}
In \figref{fig:rate_gain_net_depth} and \figref{fig:lambda_net_depth}, we observe that an \iab network with more hops has more to gain from \fd deployment. 
At low \rinr, the last-hop gain for a network with depth two is about $1.5$ whereas for a four-hop network the last-hop sum-rate improves from $300$ packets/sec to more than $2000$ packets/sec, an eight-fold rate improvement---far beyond the familiar potential doubling of capacity with \fd.
\figref{fig:lambda_per_hop} and \figref{fig:rate_gain_per_hop} present the sum-rate and rate gain across different hops for a four-hop network. 
With \fd, users at all hops throughout the network enjoy a healthy rate, unlike in \hd-\iab, where only first-hop users see a high rate (even having used a logarithmic utility).
UEs deeper in the network see higher rate gains with \fd, since their rates under \hd-\iab are so poor.

Note that the rate gain for the first hop UE is about $0.5$. 
This is because of the shape of the constraint set in \eqref{opt:max_delay} for \hd-\iab.
With the \hd restriction, and even with the logarithmic utility, the first-hop UEs are greedily served because the third- and fourth-hop UEs can meet the delay constraint with very small arrival rates.
However, a more fair operating point can be reached with \fd-\iab. 
Thus, even though it seems to be sacrificing performance at the first-hop, in reality \fd-\iab improves the overall network performance, measured by the logarithmic utility.
 
\begin{figure}
    \centering
    \subfloat[At last hop for various network depths.]{%
        \includegraphics[width=3in,
                        keepaspectratio]{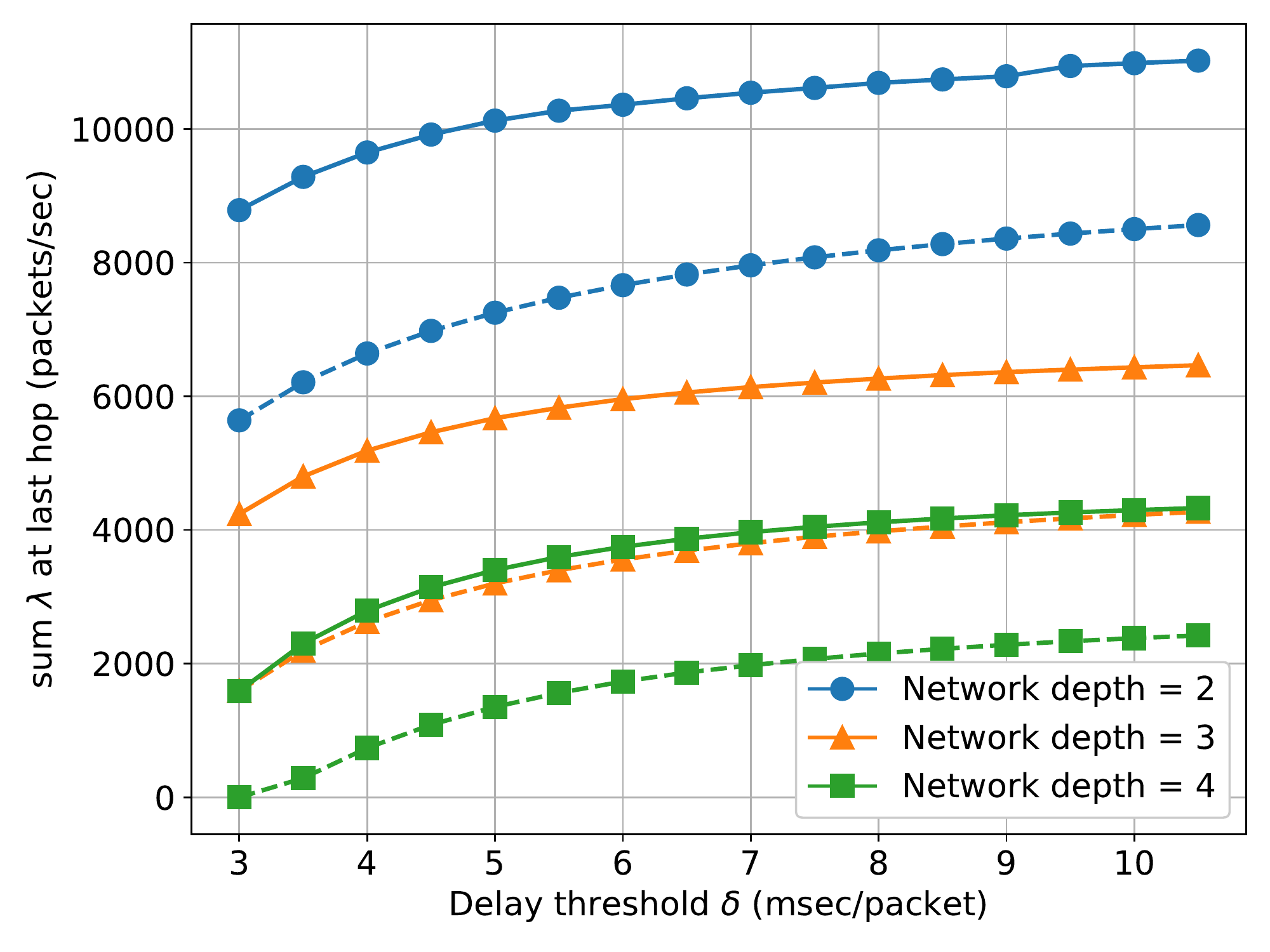}%
        \label{fig:lambda_delta_net_depth}}
    \quad
    \subfloat[At various hops for network depth $K+1=4$.]{%
        \includegraphics[width=3in,
                     keepaspectratio]{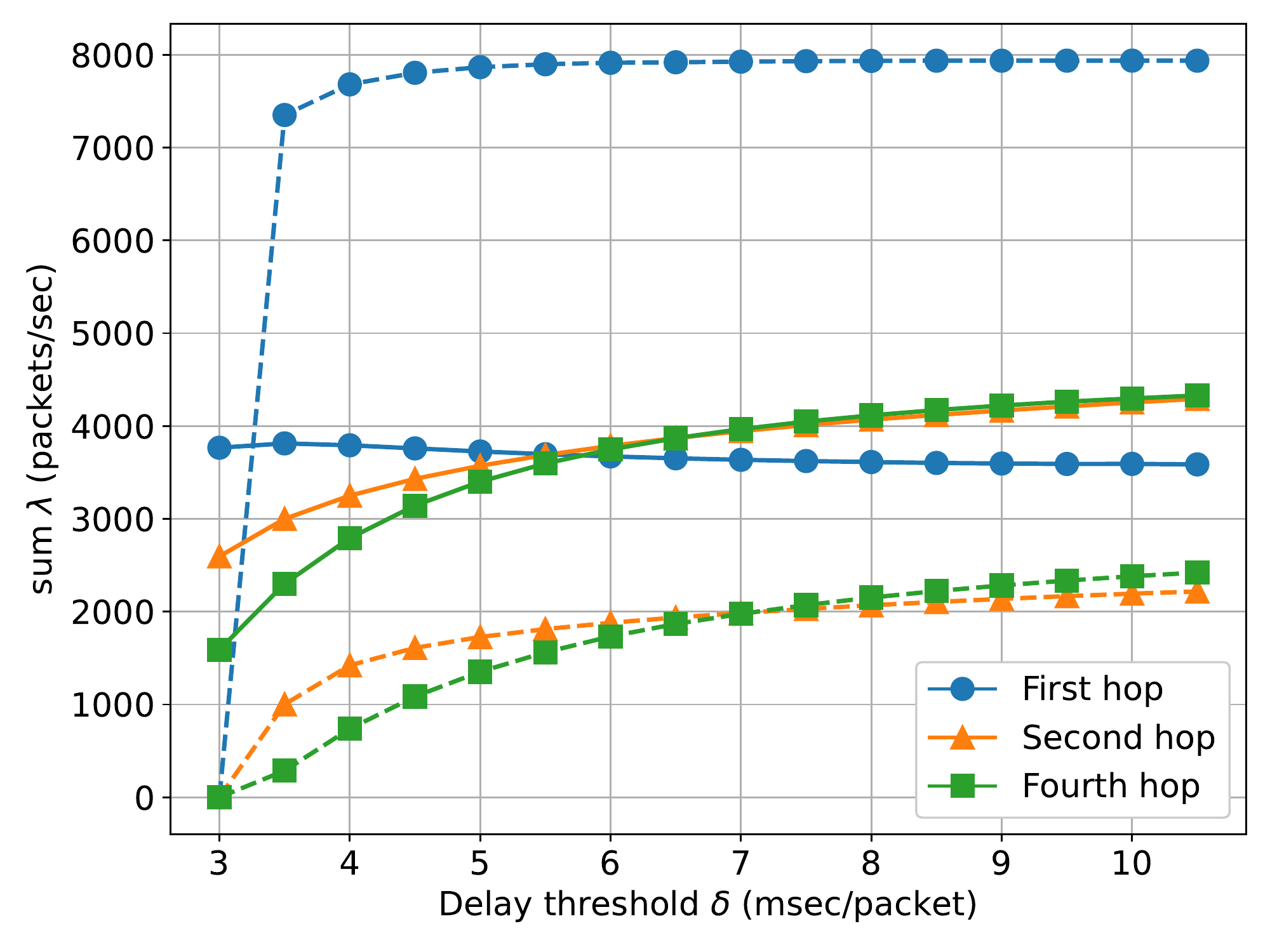}%
        \label{fig:lambda_delta_per_hop}}
    \caption{(a) Sum-rate for a last-hop user as a function of delay threshold $\delta$.
    (b) Sum-rate for users at various hops as a function of delay threshold $\delta$.
    Solid lines represent \fd-\iab and dashed lines represent the \hd-\iab
    , and for \fd we consider  $\mrinr = -15$ dB. 
    Supporting practical arrivals rates at third and fourth hops for strict delay thresholds is feasible with \fd-\iab.}
    \label{fig:lambda_delta_hd_vs_fd}
\end{figure}

\begin{figure}
    \centering
    \subfloat[At the last hop for various network depths.]{%
        \includegraphics[width=3in,keepaspectratio]{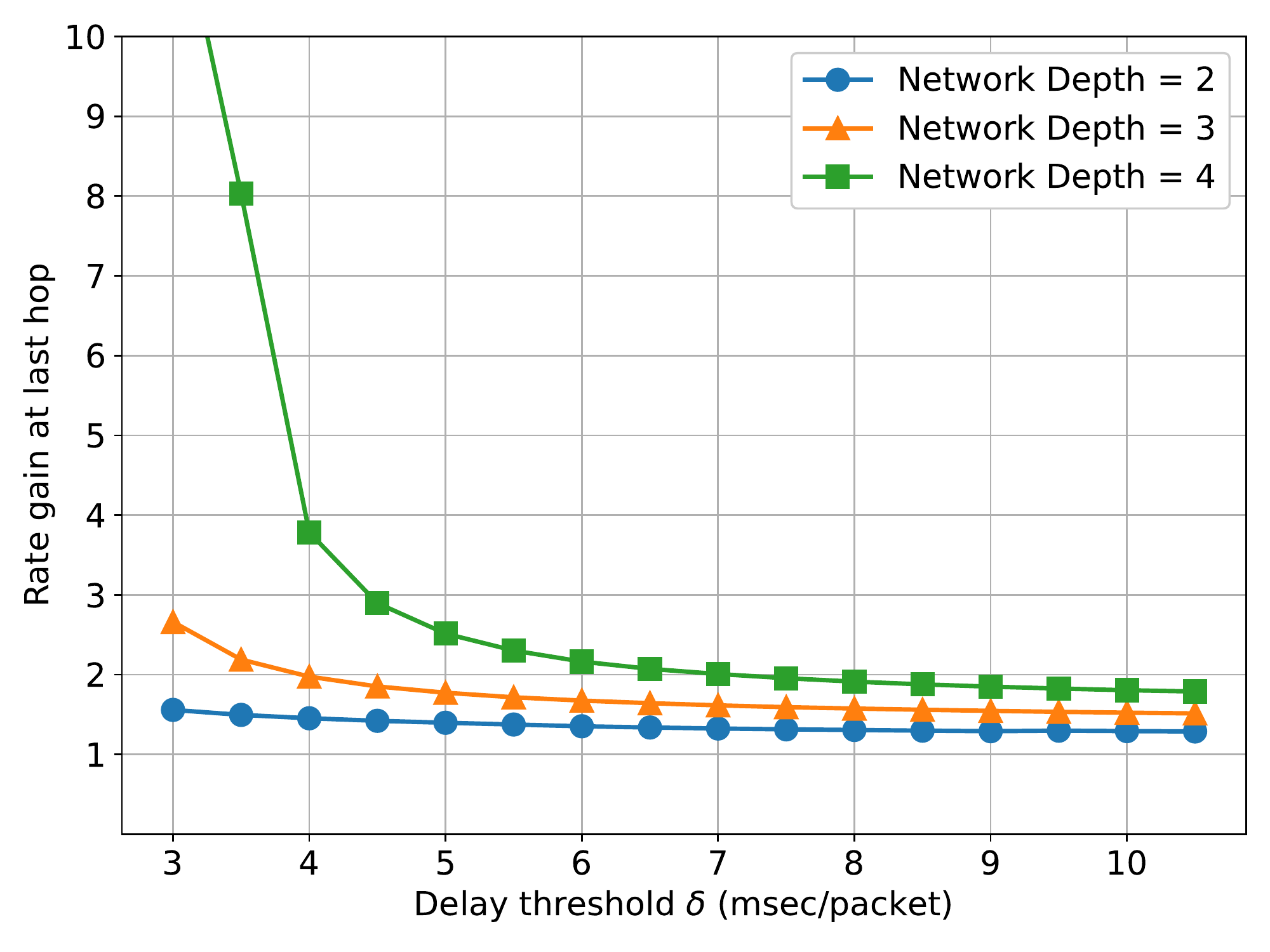}%
        \label{fig:rate_gain_delta_net_depth}}
    \quad
    \subfloat[At various hops for network depth $K+1=4$.]{%
        \includegraphics[width=3in,
                        keepaspectratio]{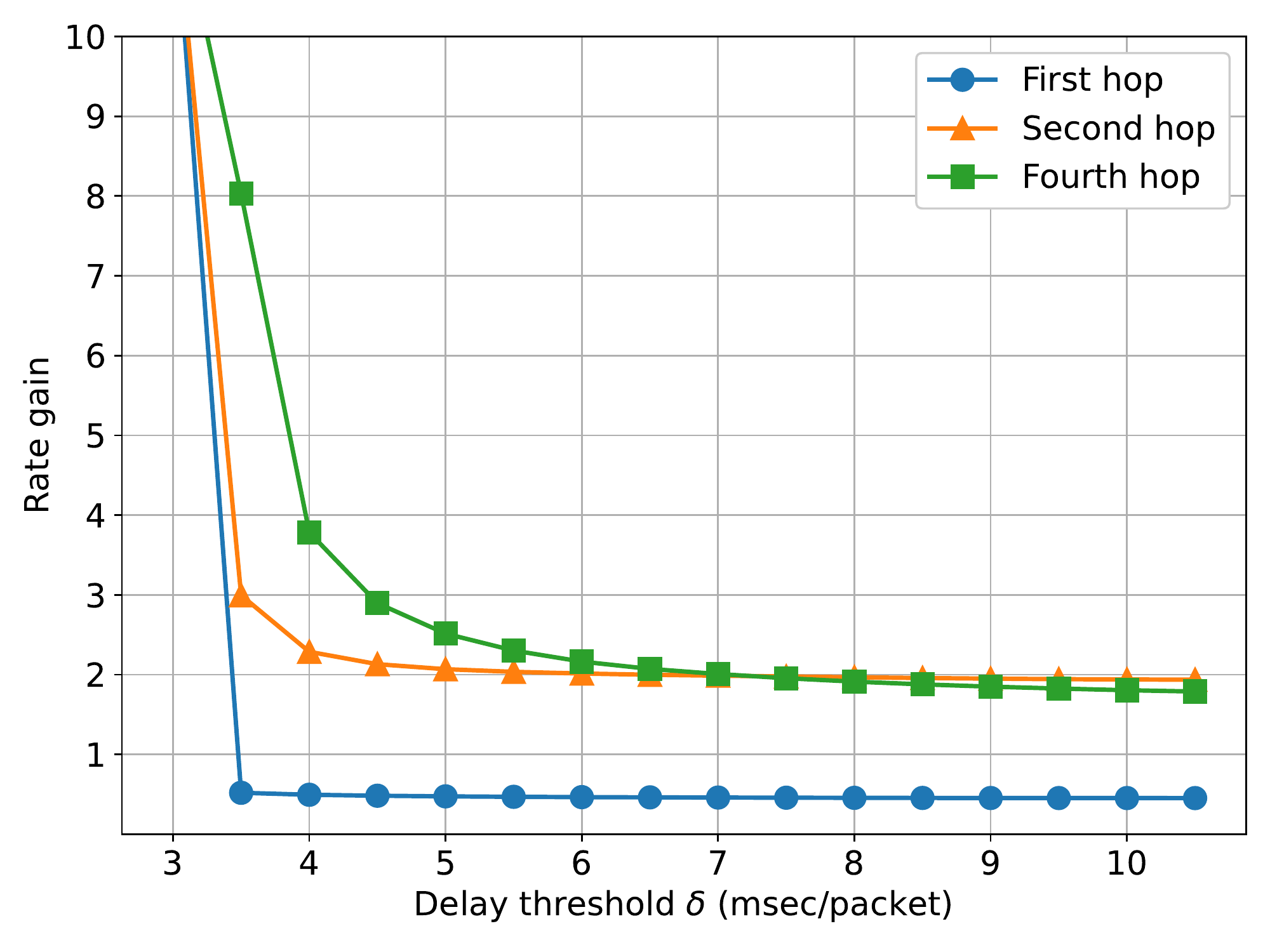}%
        \label{fig:rate_gain_delta_per_hop}}
    \caption{(a) From \figref{fig:lambda_delta_net_depth}, the rate gain for a last-hop user as a function of delay threshold $\delta$.
    (b) From \figref{fig:lambda_delta_per_hop}, the rate gain for users at various hops as a function of delay threshold $\delta$.
    As stricter delay thresholds are enforced, \fd-\iab offers increasing rate gain at the last hop over \hd-\iab; very strict delay thresholds that cannot be met by \hd-\iab can be met by \fd-\iab.}
    \label{fig:rate_gain_delta_hd_vs_fd}
\end{figure}

\textbf{\fd-\iab allows for tighter delay constraints.}
In our previous results, we fixed a delay threshold $\delta$ in examining rate gain.
Now we evaluate rate gain for changes in delay threshold $\delta$.
In \figref{fig:lambda_delta_net_depth} and \figref{fig:rate_gain_delta_net_depth}, we plot last-hop sum-rate and last-hop rate gain as a function of delay threshold $\delta$ for various network depths. 
In doing so, we consider an $\mrinr = -15$ dB in light of the diminishing gains at low \rinr, with the understanding that increased residual \si would degrade the backhaul capacities and affect the rate gain as in \figref{fig:rate_gain_hd_vs_fd}.
The rate improvement offered by \fd-\iab is seen across network depths, with deeper networks experiencing a higher rate gain as before.
By alleviating the multiplexing delay at the \iab nodes, \fd-\iab improves the arrival rates at stricter delay constraints. In fact, delay constraints which were infeasible for \hd-\iab can be made feasible by upgrading to \fd. 
Note that in \figref{fig:rate_gain_delta_net_depth}, the gain at the fourth hop for $\delta = 3$ msec/packet is infinite, because the corresponding \hd-\iab deployment is infeasible (rate of $0$ packets/sec in \figref{fig:lambda_delta_net_depth}). 
The rate gain saturates with increasing $\delta$, because for large delay thresholds the per-user arrival rate is limited by the backhaul capacities and not the multiplexing delay at the \iab node.

\figref{fig:lambda_delta_per_hop} and \figref{fig:rate_gain_delta_per_hop} show the sum-rate and rate gain across different hops for a four-hop network. 
The observations are consistent with the story so far: UEs further from the donor have more to benefit from \fd-\iab. 
At very small $\delta$, the gain tends to infinity since a delay constraint that cannot be met by \hd-\iab can in fact be met with \fd.
Like before, we see \fd-\iab is capable of delivering fairer service to UEs throughout the network, while still meeting the delay constraints across multiple hops.
\begin{figure}[t!]
    \centering
    \subfloat[Minimum feasible delay threshold.]{%
        \includegraphics[width=3in,keepaspectratio]{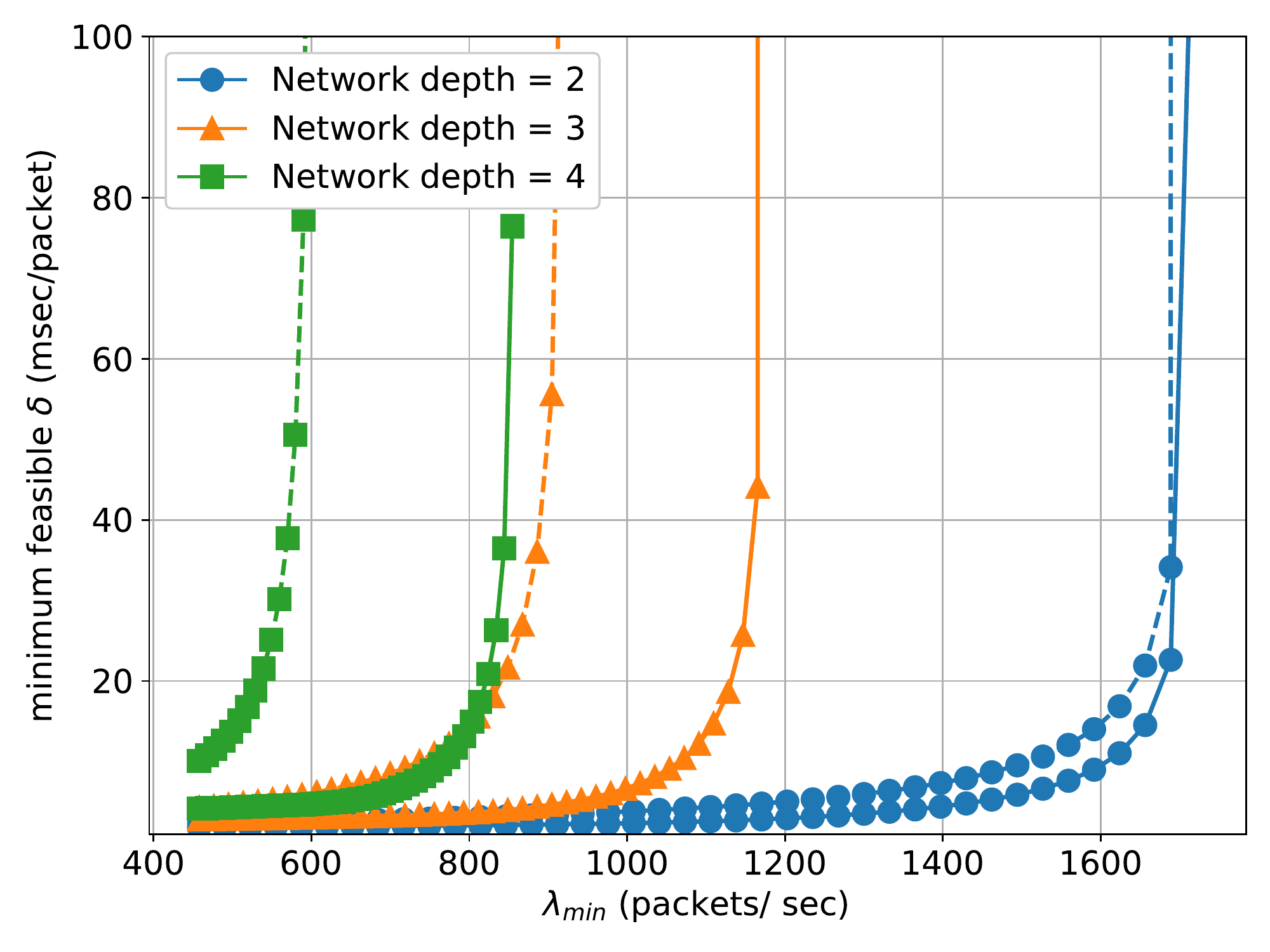}%
        \label{fig:delta_vs_minlam}}
    \quad
    \subfloat[Latency gain $\ell$.]{%
        \includegraphics[width=3in,keepaspectratio]{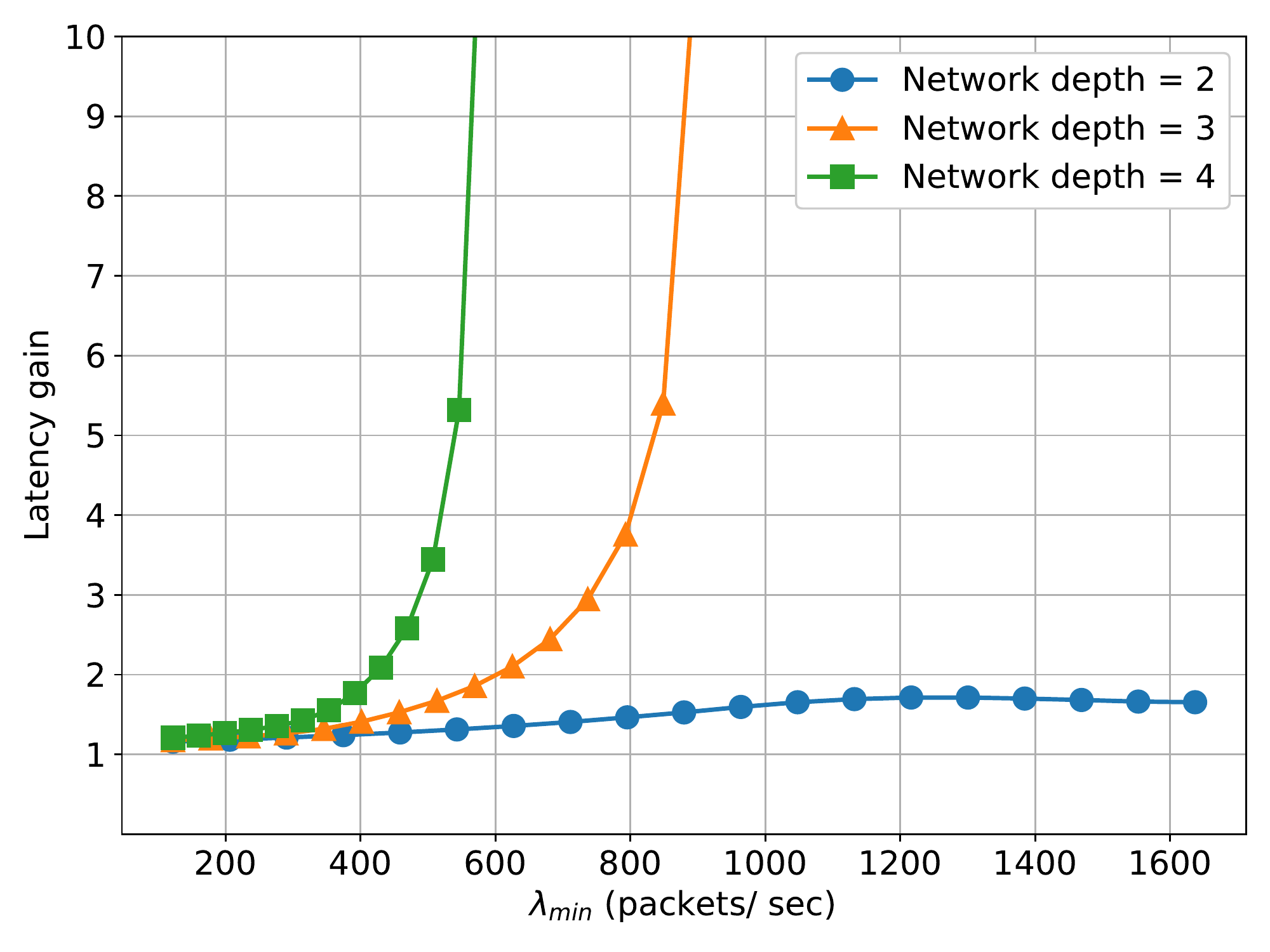}%
        \label{fig:latency_gain}}
    \caption{(a) Minimum feasible delay $\delta^*$ of \fd-\iab (solid) and \hd-\iab (dashed) as a function of minimum rate requirement $\minlam$ for various network depths, where $\mrinr = -15$ dB. (b) From (a), the corresponding latency gain $\ell$ of \fd-\iab over \hd-\iab as a function of minimum rate requirement $\minlam$ for various network depths.}
    \label{fig:latency_hd_vs_fd}
\end{figure} 

\textbf{\fd-\iab supports higher per-UE rates.} 
To study improvements in per-UE rate with delay constraint, we solve the linear program \eqref{opt:delay_feas_lp_reform} for both \fd-\iab and \hd-\iab modes and employ our definition of latency gain \eqref{eq:latency_gain_general}. 
We perform simulations fixing $\mrinr=-15$ dB and examine the $(\minlam,\delta^*)$-pairs that can be met by \hd and those that can be met by \fd.
\figref{fig:latency_gain} illustrates these results for various network depths, where curves offering higher $\minlam$ at lower $\delta^*$ are more desirable (i.e., toward the lower right).
For a two-hop network, both the \hd and the \fd deployments handle the latency and throughput constraints equally well.
With more than two hops, however, the power of \fd in this network setting is brought to light.
For a given latency target, \fd-\iab can deliver a minimum rate $\minlam$ far greater than \hd-\iab: for $\delta^* = 20$ msec/packet, \fd offers over $1.5$ times higher $\minlam$ in a four-hop network and over $1.3$ times higher $\minlam$ in a three-hop network.
Minimum rate targets $\minlam$ that can be met by \hd-\iab networks only with prohibitively high delays, can also be met by \fd-\iab while simultaneously offering low latency.
For instance, in a three-hop network, \hd-\iab can deliver $\minlam = 900$ packets/sec only be introducing delays well beyond what is practical; a \fd-\iab network can do so while meeting a latency of approximately $\delta^* = 5$ msec/packet.
As a result, latency gain can tend toward infinity.


\subsection{Beyond Line Networks}

\begin{figure}[t]
    \centering
    \includegraphics[height=2.5in,
                    keepaspectratio]{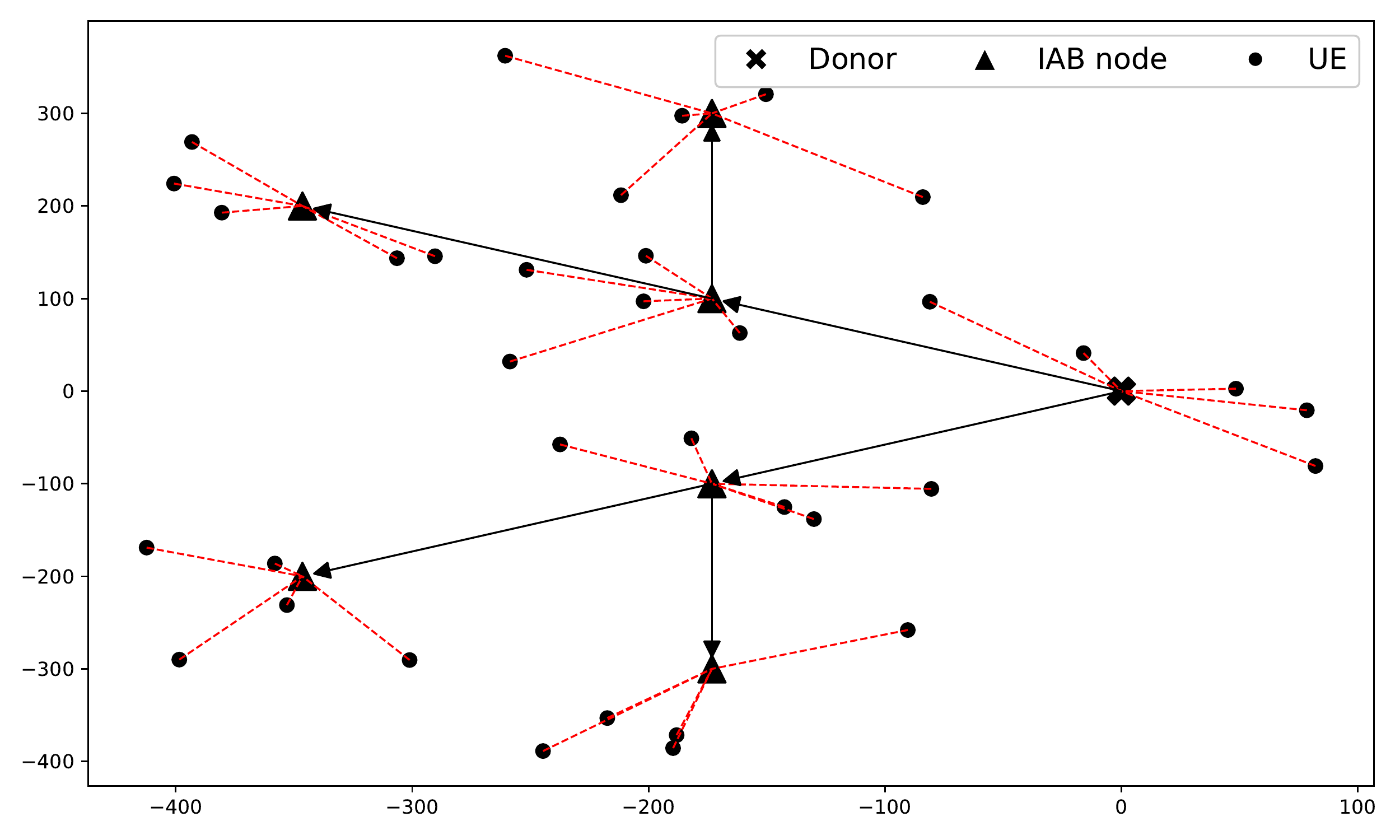}
    \caption{A realization of the simulated two-child tree network with one donor and six \iab nodes. The donor and each \iab serves five UEs. Note that first-hop \iab nodes deliver backhaul to child \iab nodes, along with serving access.}
    \label{fig:symmetric_tree}
\end{figure}

While line networks are expected to be common for initial \iab deployments, the presented framework is in no way limited to them. 
To briefly illustrate this, we use our framework to study the \iab deployment in \figref{fig:symmetric_tree}, which consists of one donor node, $K=6$ \iab nodes, and five UEs per BS.
We refer to this as a \textit{two-child tree network} since the donor and each first-hop \iab node deliver backhaul to two \iab nodes.
While not explicitly shown due to space constraints, the key takeaways from the discussion on line networks holds, namely: 
(i) rate gain saturates for \rinr below around $-5$ dB, 
(ii) UEs further from the donor have more to benefit from an \fd-\iab deployment, and
(iii) \fd-\iab can support larger $\minlam$ for the same minimum feasible delay $\delta^*$ as was shown in \figref{fig:latency_hd_vs_fd}.

\begin{figure}[t!]
    \centering
        \subfloat[Mean per-user arrival rate.]{%
        \includegraphics[width=3in,keepaspectratio]{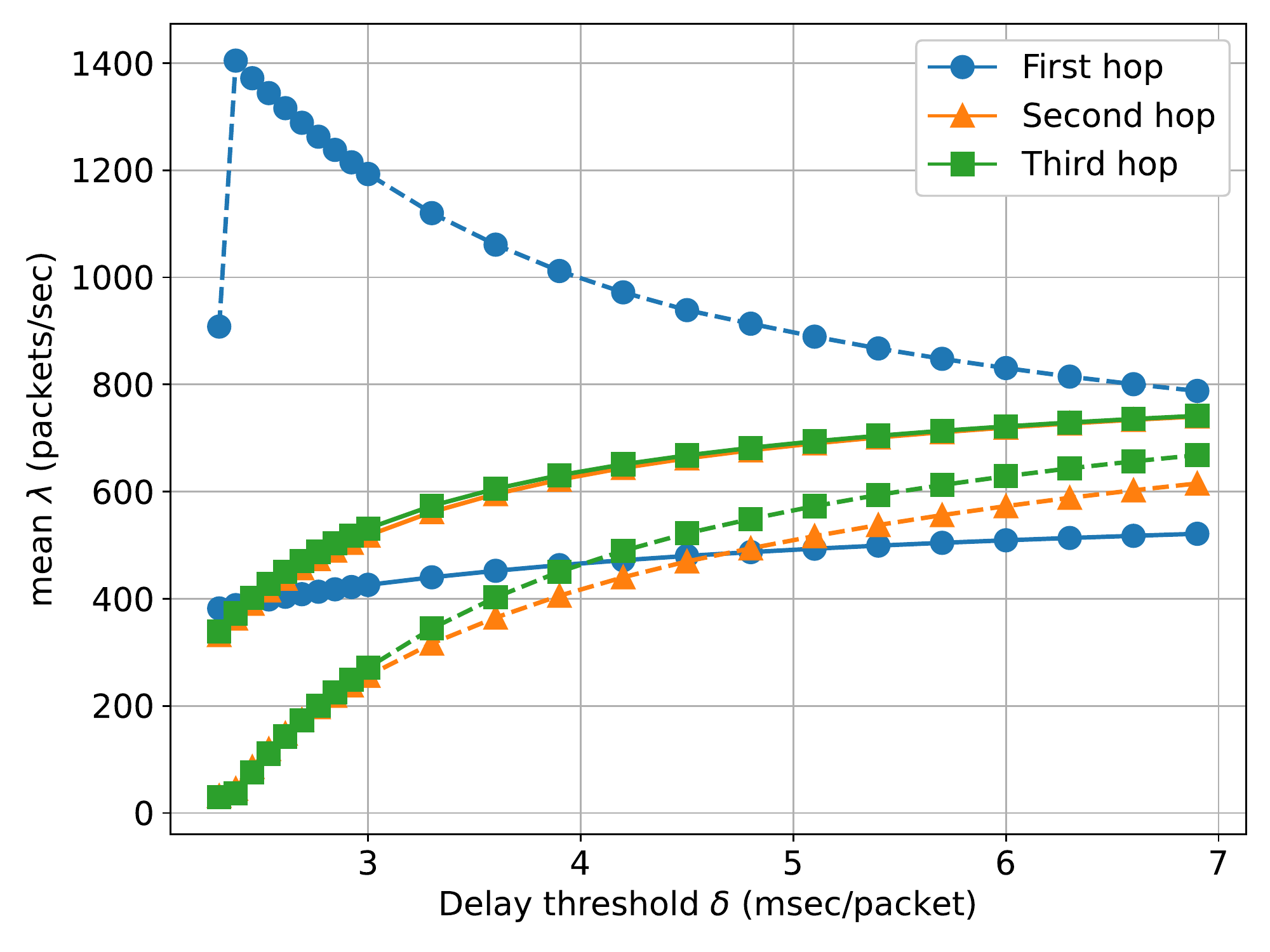}%
        \label{fig:lambda_vs_delta_tree1}}
        \quad
        \subfloat[Rate gain.]{%
        \includegraphics[width=3in,keepaspectratio]{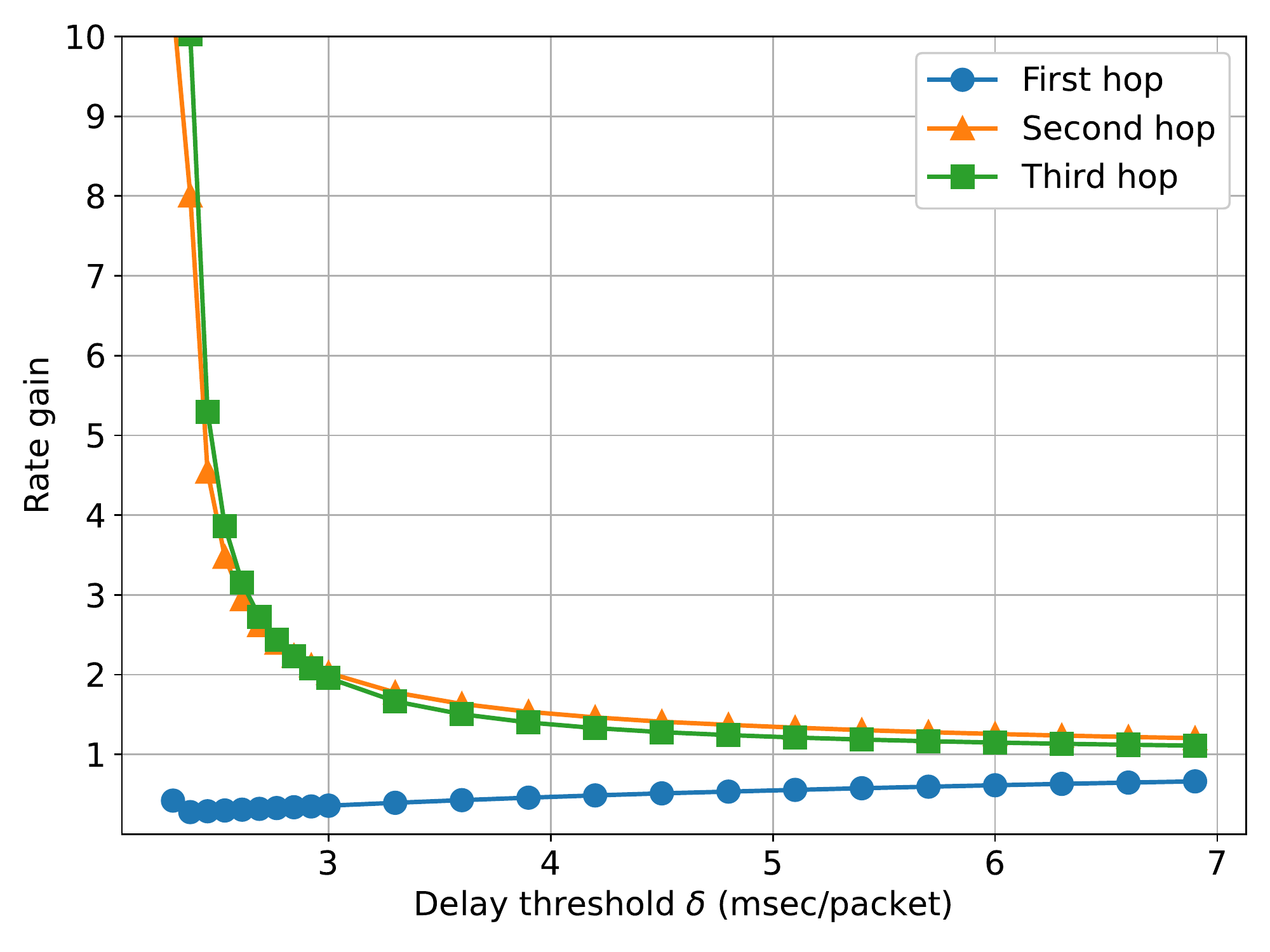}%
        \label{fig:gain_vs_delta_tree1}}
        \caption{(a) Mean per-UE rate of \fd-\iab (solid) and \hd-\iab (dashed), and (b) Rate gain of \fd-\iab over \hd-\iab, as a function of delay threshold $\delta$ at various hops in the two-child network in \figref{fig:symmetric_tree}, where $\mrinr = -15$ dB.
        \fd-\iab alleviates the resource bottleneck due to branching at the donor and achieves a fairer rate distribution.}
        \label{fig:lambda_tree}
\end{figure} 

Nonetheless, there are some additional insights to draw from this two-child tree network.
To start, we consider \figref{fig:lambda_vs_delta_tree1}, which shows the mean per-user arrival rate at various hops in the network as a function delay threshold.
Unlike line networks, the donor and first-hop \iab nodes now deliver backhaul to more than one \iab node. 
Consequently, a \hd-\iab deployment cannot support strict latency requirements at the second- and third-hop with a practically viable arrival rate.
This is because first-hop \iab nodes must multiplex incoming and outgoing backhaul links, significantly worsening the multiplexing delay and the radio resource bottleneck.
First-hop users, however, can enjoy high rates---even under strict delay thresholds---since they are not subjected to these relaying delays.
The disparity in user service is evident in \figref{fig:lambda_vs_delta_tree1} as the delay threshold is made stricter, with first-hop arrival rate increasing as users deeper in the network see their service degrading.



With \fd-\iab, however, the resource bottleneck is alleviated and the network can serve its multihop users much more fairly and with satisfactory rates.
We can see from \figref{fig:lambda_vs_delta_tree1} that \fd-\iab can support delay thresholds infeasible for its \hd counterpart, with a healthy throughput across hops.
The corresponding rate gain of \fd-\iab over \hd-\iab is shown in \figref{fig:gain_vs_delta_tree1}.
\fd-\iab can meet very strict delay thresholds for second- and third-hop users when \hd-\iab cannot, driving the rate gain toward infinity as $\delta$ is decreased.
As $\delta$ increases, the rate gain saturates like in \figref{fig:rate_gain_delta_per_hop}.
For first-hop users, we observe a rate gain less than one, which may seem undesirable at first glance but like \figref{fig:rate_gain_per_hop} is justified by \fd-\iab achieving a fairer distribution of arrival rates across hops.

Comparing these results of a two-child tree network to that of the simulated line network, we notice a few key differences.
First, comparing \figref{fig:gain_vs_delta_tree1} to \figref{fig:rate_gain_delta_per_hop}, we see that the rate gain of the second- and third-hop users in the two-child tree network saturates to noticeably less than $2$, while that for the line network saturates to nearly $2$. 
This is difficult to explain precisely, though it is likely largely due to the fact that the donor and each first-hop \iab node must multiplex outgoing backhaul transmissions---which line networks do not suffer from.
\fd cannot directly alleviate these multiplexing costs, suggesting that the network may be bottlenecked by the resources consumed when juggling multiple backhual links.
This motivates future work to more thoroughly study this disparity across network topologies and investigate how to best deploy \iab networks when \iab nodes are \fd-equipped. 
\section{Conclusion}
\label{Sec:Conc}

We present a framework for the throughput and latency of \iab networks, which can be readily used by network engineers to evaluate and optimize \iab deployments.
We use this framework to study the merits of equipping \iab nodes with \fd capability, as a means to alleviate resource bottlenecks traditionally faced by \iab networks.
Analytical and numerical results illustrate the network performance improvements of \fd-\iab over \hd-\iab and show that the former offers a multitude of benefits.
\fd-\iab can facilitate lower latency, higher throughput, deeper networks, and fairer service compared to conventional \hd-\iab. 
\fd upgrades can be particularly transformative for deeper networks, where users at or near the last hop---which may suffer in \hd-\iab deployments---can enjoy improvements in rate and latency that dramatically enhance their quality-of-service.
Furthermore, \fd-\iab widens the feasible operating region of the network, allowing it to support latency and rate targets that \hd-\iab fundamentally cannot meet.
We show that these gains can be observed in the presence of residual \si that is near or even above the noise floor.
These results motivate the use of \fd to alleviate the obstacles in latency and rate scaling suffered by traditional multihop \iab networks.

\appendices
\section{Proof of \thmref{thm:feasible_delay_sol_ra_rb}}
\label{app:feasible_delay_sol_ra_rb}
\begin{proof}
Note that, $\left[\bG_{\hdrm}\right]_{k,l}=1$ if vertex $l$ is the parent of $k$ where as $\left[\bG_{\fdrm}\right]_{k,l}=0$. 
We define
\begin{align*}
f_{\hdrm}(k) = \frac{1 - \minlam(\bG_{\hdrm})_{k,:}\bC^{-1}\bF\bm{1}}{(\bG_{\hdrm})_{k,:}\bC^{-1}\tilde{\bh}} \quad
f_{\fdrm}(k) = \frac{1 - \minlam(\bG_{\fdrm})_{k,:}\bC^{-1}\bF\bm{1}}{(\bG_{\fdrm})_{k,:}\bC^{-1}\tilde{\bh}}    
\end{align*}

\textit{The half-duplex case.}
Since we know the structure of the scheduling matrices, we have
\begin{align*}
    f_{\hdrm}(k) = \begin{cases}
                    \frac{1 - w\minlam\left(\frac{1}{R_a} + \frac{K}{R_b}\right)}{\frac{K+1}{R_b} + \frac{w}{R_a}},~k=0\\
                    \frac{1 - w\minlam\left(\frac{1}{R_a} + \frac{2(K-k)+1}{R_b}\right)}{\frac{2(K+1)}{R_b} + \frac{w(k+1)}{R_a}},~1\leq k < K \\
                    \frac{1 - w\minlam\left(\frac{1}{R_a} + \frac{1}{R_b}\right)}{\frac{K+1}{R_b} + \frac{w(K+1)}{R_a}},~k=K
                    \end{cases}.
\end{align*}
$f_{\hdrm}(1)$ has a smaller numerator and larger denominator compared to $f_{\hdrm}(0)$, and therefore $f_{\hdrm}(0) \geq f_{\hdrm}(1)$. 
Similarly, $f_{\hdrm}(K) \geq f_{\hdrm}(K-1)$. 
Hence, $k=0$ and $k=K$ will never result in the bottleneck inequality. 
By observing the derivative of $f_{\hdrm}(k)$ for $1\leq k < K$, 
\begin{align*}
    \frac{df_{\hdrm}(k)}{dk} = \frac{w\minlam\left(\frac{2w(k+1)}{R_aR_b} + \frac{4(K+1)}{R_b^2} + \frac{w(2(K-k)+1)}{R_aR_b} + \frac{w}{R^2_a}\right) - \frac{w}{R_a}}{\left(\frac{2(K+1)}{R_b} + \frac{w(k+1)}{R_a}\right)^2}, ~1\leq k < K.
\end{align*}
we see that $f_{\hdrm}(k)$ is decreasing if $\minlam \leq \frac{R_a}{4(K+1)\left(\frac{R_a}{R_b}\right)^2 + (2K+3)w\left(\frac{R_a}{R_b}\right) + w}$
and increasing otherwise. 
Therefore, for the decreasing case, the bottleneck inequality is $f_{\hdrm}(K-1)$ and for the increasing case, the bottleneck inequality is $f_{\hdrm}(1)$.

\textit{The full-duplex case.}
Starting similarly with $f_{\fdrm}(k)$, we have
\begin{align*}
    f_{\fdrm}(k) = \begin{cases}
                    \frac{1 - w\minlam\left(\frac{1}{R_a} + \frac{K}{R_b}\right)}{\frac{K+1}{R_b} + \frac{w}{R_a}},~k=0\\
                    \frac{1 - w\minlam\left(\frac{1}{R_a} + \frac{K-k}{R_b}\right)}{\frac{K+1}{R_b} + \frac{w(k+1)}{R_a}},~1\leq k < K \\
                    \frac{1 - \frac{w\minlam}{R_a}}{ \frac{w(K+1)}{R_a}},~k=K
                    \end{cases}.
\end{align*}
Note that $f_{\fdrm}(K) \geq f_{\fdrm}(K-1)$, allowing us to draw analogous conclusions from the \hd case.
Taking the derivative of $f_{\fdrm}(k)$ for $1\leq k < K$, 
\begin{align*}
    \frac{df_{\fdrm}(k)}{dk} = \frac{w\minlam\left(\frac{w(k+1)}{R_aR_b} + \frac{K+1}{R_b^2} + \frac{w(K-k)}{R_aR_b} + \frac{w}{R^2_a}\right) - \frac{w}{R_a}}{\left(\frac{K+1}{R_b} + \frac{w(k+1)}{R_a}\right)^2}, ~1\leq k < K
\end{align*}
we see that when $\minlam<\frac{R_a}{(K+1)\left(\frac{R_a}{R_b}\right)^2 + w(K+1)\left(\frac{R_a}{R_b}\right) + w}$, 
$f_{\fdrm}(k)$ is decreasing (over $1\leq k < K$) and $f_{\fdrm}(0) > f_{\fdrm}(1)$. Hence, the bottleneck inequality is $f_{\fdrm}(K-1)$. 
On the other hand, if $\minlam\geq\frac{R_a}{(K+1)\left(\frac{R_a}{R_b}\right)^2 + w(K+1)\left(\frac{R_a}{R_b}\right) + w}$,
then $f_{\fdrm}(k)$ is increasing for $1\leq k < K$ and $f_{\fdrm}(0) \leq f_{\fdrm}(1)$. 
Hence, the bottleneck inequality is $f_{\fdrm}(0)$.
\end{proof}

\bibliographystyle{IEEEtran}
\bibliography{MananRef}

\begin{thebibliography}{10}
\providecommand{\url}[1]{#1}
\csname url@samestyle\endcsname
\providecommand{\newblock}{\relax}
\providecommand{\bibinfo}[2]{#2}
\providecommand{\BIBentrySTDinterwordspacing}{\spaceskip=0pt\relax}
\providecommand{\BIBentryALTinterwordstretchfactor}{4}
\providecommand{\BIBentryALTinterwordspacing}{\spaceskip=\fontdimen2\font plus
\BIBentryALTinterwordstretchfactor\fontdimen3\font minus
  \fontdimen4\font\relax}
\providecommand{\BIBforeignlanguage}[2]{{%
\expandafter\ifx\csname l@#1\endcsname\relax
\typeout{** WARNING: IEEEtran.bst: No hyphenation pattern has been}%
\typeout{** loaded for the language `#1'. Using the pattern for}%
\typeout{** the default language instead.}%
\else
\language=\csname l@#1\endcsname
\fi
#2}}
\providecommand{\BIBdecl}{\relax}
\BIBdecl

\bibitem{Pi11}
Z.~Pi and F.~Khan, ``An introduction to millimeter-wave mobile broadband
  systems,'' \emph{{IEEE} Commun. Mag.}, vol.~49, no.~6, pp. 101--107, June
  2011.

\bibitem{Millimeter_Rappaport13}
T.~S. Rappaport \emph{et~al.}, ``Millimeter wave mobile communications for {5G}
  cellular: {I}t will work!'' \emph{IEEE Access}, vol.~1, no.~1, pp. 335--349,
  May 2013.

\bibitem{heath16overview}
R.~W. Heath, N.~González-Prelcic, S.~Rangan, W.~Roh, and A.~M. Sayeed, ``An
  overview of signal processing techniques for millimeter wave {MIMO}
  systems,'' \emph{{IEEE} J. Sel. Topics Signal Process.}, vol.~10, no.~3, pp.
  436--453, Apr 2016.

\bibitem{rangan2014millimeter}
S.~Rangan, T.~S. Rappaport, and E.~Erkip, ``Millimeter-wave cellular wireless
  networks: Potentials and challenges,'' \emph{Proc. {IEEE}}, vol. 102, no.~3,
  pp. 366--385, Mar. 2014.

\bibitem{3gpp_tr38874}
{3GPP TR 38.874}, in \emph{Study on integrated access and backhaul},
  \url{http://ftp.3gpp.org//Specs/archive/38\_ series/38.874/38874-100.zip},
  Aug. 2020.

\bibitem{cudak21integrated}
M.~Cudak, A.~Ghosh, A.~Ghosh, and J.~G. Andrews, ``Integrated access and
  backhaul: A key enabler for {5G} millimeter-wave deployments,'' \emph{{IEEE}
  Commun. Mag.}, vol.~59, no.~4, pp. 88--94, April 2021.

\bibitem{rasek20joint}
M.~Eslami~Rasekh, D.~Guo, and U.~Madhow, ``Joint routing and resource
  allocation for millimeter wave picocellular backhaul,'' \emph{{IEEE} Trans.
  Wireless Commun.}, vol.~19, no.~2, pp. 783--794, Feb. 2020.

\bibitem{gupta2020andrews}
M.~Gupta, A.~Rao, E.~Visotsky, A.~Ghosh, and J.~G. Andrews, ``Learning link
  schedules in self-backhauled millimeter wave cellular networks,''
  \emph{{IEEE} Trans. Wireless Commun.}, vol.~19, no.~12, pp. 8024--8038, Dec.
  2020.

\bibitem{ortiz19scaros}
A.~Ortiz, A.~Asadi, G.~H. Sim, D.~Steinmetzer, and M.~Hollick, ``Scaros: A
  scalable and robust self-backhauling solution for highly dynamic
  millimeter-wave networks,'' \emph{{IEEE} J. Sel. Areas Commun.}, vol.~37,
  no.~12, pp. 2685--2698, Dec 2019.

\bibitem{du17gbps}
J.~Du, E.~Onaran, D.~Chizhik, S.~Venkatesan, and R.~A. Valenzuela, ``Gbps user
  rates using mm{W}ave relayed backhaul with high gain antennas,'' \emph{{IEEE}
  J. Sel. Areas Commun.}, vol.~35, no.~6, pp. 1363--1372, June 2017.

\bibitem{nazmul17sampath}
M.~N. Islam, S.~Subramanian, and A.~Sampath, ``Integrated access backhaul in
  millimeter wave networks,'' in \emph{Proc. IEEE WCNC}, April 2017.

\bibitem{kulkarni17performance}
M.~N. Kulkarni, J.~G. Andrews, and A.~Ghosh, ``Performance of dynamic and
  static {TDD} in self-backhauled millimeter wave cellular networks,''
  \emph{{IEEE} Trans. Wireless Commun.}, vol.~16, no.~10, pp. 6460--6478, Oct
  2017.

\bibitem{jain2005impact}
K.~Jain, J.~Padhye, V.~N. Padmanabhan, and L.~Qiu, ``Impact of interference on
  multi-hop wireless network performance,'' \emph{Wireless Networks}, vol.~11,
  no.~4, pp. 471--487, Jul 2005.

\bibitem{gupta00capacity}
P.~Gupta and P.~R. Kumar, ``The capacity of wireless networks,'' \emph{{IEEE}
  Trans. Inf. Theory}, vol.~46, no.~2, pp. 388--404, March 2000.

\bibitem{fra09capaccity}
M.~Franceschetti, M.~D. Migliore, and P.~Minero, ``The capacity of wireless
  networks: Information-theoretic and physical limits,'' \emph{IEEE Trans. Inf.
  Theory}, vol.~55, no.~8, pp. 3413--3424, Aug 2009.

\bibitem{zemlianov05capacity}
A.~Zemlianov and G.~de~Veciana, ``Capacity of ad hoc wireless networks with
  infrastructure support,'' \emph{{IEEE} J. Sel. Areas Commun.}, vol.~23,
  no.~3, pp. 657--667, March 2005.

\bibitem{polese2020integrated}
M.~Polese, M.~Giordani, T.~Zugno, A.~Roy, S.~Goyal, D.~Castor, and M.~Zorzi,
  ``Integrated access and backhaul in 5{G} {mmWave} networks: Potential and
  challenges,'' \emph{{IEEE} Commun. Mag.}, vol.~58, no.~3, pp. 62--68, Mar.
  2020.

\bibitem{polese2018distributed}
M.~Polese, M.~Giordani, A.~Roy, D.~Castor, and M.~Zorzi, ``Distributed path
  selection strategies for integrated access and backhaul at {mmWaves},'' in
  \emph{Proc. IEEE GLOBECOM}, Dec. 2018, pp. 1--7.

\bibitem{vu19joint}
T.~K. Vu, M.~Bennis, M.~Debbah, and M.~Latva-Aho, ``Joint path selection and
  rate allocation framework for {5G} self-backhauled mm-wave networks,''
  \emph{{IEEE} J. Sel. Areas Commun.}, vol.~18, no.~4, pp. 2431--2445, Apr
  2019.

\bibitem{cuba2020twice}
F.~Gómez-Cuba and M.~Zorzi, ``Twice simulated annealing resource allocation
  for {mmWave} multi-hop networks with interference.'' in \emph{Proc. IEEE
  ICC}, July 2020, pp. 1--7.

\bibitem{saha19millimeter}
C.~Saha and H.~S. Dhillon, ``Millimeter wave integrated access and backhaul in
  {5G}: Performance analysis and design insights,'' \emph{{IEEE} J. Sel. Areas
  Commun.}, vol.~37, no.~12, pp. 2669--2684, 2019.

\bibitem{madapatha2021topology}
C.~Madapatha, B.~Makki, A.~Muhammad, E.~Dahlman, M.-S. Alouini, and
  T.~Svensson, ``On topology optimization and routing in integrated access and
  backhaul networks: A genetic algorithm-based approach,'' \emph{{IEEE} Open J.
  Commun. Soc.}, vol.~2, pp. 2273--2291, Sept. 2021.

\bibitem{zhang2021survey}
Y.~Zhang, M.~A. Kishk, and M.-S. Alouini, ``A survey on integrated access and
  backhaul networks,'' \emph{Frontier Commun. Netw.}, vol.~2, pp. 1--24, June
  2021.

\bibitem{roberts20equipping}
I.~P. Roberts, H.~B. Jain, and S.~Vishwanath, ``Equipping millimeter-wave
  full-duplex with analog self-interference cancellation,'' in \emph{Proc. IEEE
  ICC WKSHP}, July 2020, pp. 1--6.

\bibitem{bishnu21performance}
A.~Bishnu, M.~Holm, and T.~Ratnarajah, ``Performance evaluation of full-duplex
  {IAB} multi-cell and multi-user network for {FR2} band,'' \emph{IEEE Access},
  vol.~9, pp. 72\,269--72\,283, May 2021.

\bibitem{roberts21millimeter}
I.~P. Roberts, J.~G. Andrews, H.~B. Jain, and S.~Vishwanath, ``Millimeter-wave
  full duplex radios: New challenges and techniques,'' \emph{{IEEE} Wireless
  Commun.}, vol.~28, no.~1, pp. 36--43, Feb. 2021.

\bibitem{xiao17fullduplex}
Z.~Xiao, P.~Xia, and X.-G. Xia, ``Full-duplex millimeter-wave communication,''
  \emph{{IEEE} Wireless Commun.}, vol.~24, no.~6, pp. 136--143, 2017.

\bibitem{suk21fullduplex}
\BIBentryALTinterwordspacing
G.~Y. Suk, S.-M. Kim, J.~Kwak, S.~Hur, E.~Kim, and C.-B. Chae, ``Full duplex
  integrated access and backhaul for {5G NR}: Analyses and prototype
  measurements,'' \emph{CoRR}, vol. abs/2007.03272, June 2021. [Online].
  Available: \url{https://arxiv.org/abs/2007.03272}
\BIBentrySTDinterwordspacing

\bibitem{bai2015coverage}
T.~Bai and R.~W. Heath, ``Coverage and rate analysis for millimeter-wave
  cellular networks,'' \emph{{IEEE} Trans. Wireless Commun.}, vol.~14, no.~2,
  pp. 1100--1114, Feb 2015.

\bibitem{yuan2018optimal}
D.~Yuan, H.-Y. Lin, J.~Widmer, and M.~Hollick, ``Optimal joint routing and
  scheduling in millimeter-wave cellular networks,'' in \emph{Proc. IEEE
  INFOCOM}, 2018, pp. 1205--1213.

\bibitem{kulkarni19maxmin}
\BIBentryALTinterwordspacing
M.~N. Kulkarni, A.~Ghosh, and J.~G. Andrews, ``Max-min rates in self-backhauled
  millimeter wave cellular networks,'' Aug. 2018. [Online]. Available:
  \url{https://arxiv.org/abs/1805.01040}
\BIBentrySTDinterwordspacing

\bibitem{srikant2014comnets}
R.~Srikant and L.~Ying, \emph{Communication Networks: An Optimization, Control
  and Stochastic Networks Perspective}.\hskip 1em plus 0.5em minus 0.4em\relax
  New York, NY, USA: Cambridge University Press, 2014.

\bibitem{3gpp_tr36889}
{3GPP TR 38.889}, in \emph{Feasibility Study on Licensed-Assisted Access to
  Unlicensed Spectrum},
  \url{http://www.3gpp.org/ftp/Specs/archive/36\_series/36.889/36889-d00.zip},
  July 2015.

\bibitem{yigal07fairness}
Y.~Bejerano, S.-J. Han, and L.~Li, ``Fairness and load balancing in wireless
  {LANs} using association control,'' \emph{IEEE/ACM Trans. on Netw.}, vol.~15,
  no.~3, pp. 560--573, June 2007.

\bibitem{ye13user}
Q.~{Ye}, B.~{Rong}, Y.~{Chen}, M.~{Al-Shalash}, C.~{Caramanis}, and J.~G.
  {Andrews}, ``User association for load balancing in heterogeneous cellular
  networks,'' \emph{{IEEE} Trans. Wireless Commun.}, vol.~12, no.~6, pp.
  2706--2716, June 2013.

\bibitem{kim12distributed}
H.~Kim, G.~de~Veciana, X.~Yang, and M.~Venkatachalam, ``Distributed
  $\alpha$-optimal user association and cell load balancing in wireless
  networks,'' \emph{IEEE/ACM Trans. on Netw.}, vol.~20, no.~1, pp. 177--190,
  Feb. 2012.

\bibitem{fettweis2014tactile}
G.~P. Fettweis, ``The tactile internet: Applications and challenges,''
  \emph{{IEEE} Veh. Technol. Mag.}, vol.~9, no.~1, pp. 64--70, Mar. 2014.

\bibitem{mfm_arxiv}
\BIBentryALTinterwordspacing
I.~P. Roberts, ``{MIMO} for {MATLAB}: A toolbox for simulating {MIMO}
  communication systems,'' Nov. 2021. [Online]. Available:
  \url{https://arxiv.org/abs/2111.05273}
\BIBentrySTDinterwordspacing

\bibitem{3gpp_tr_38901}
{3GPP TR 38.901}, in \emph{Study on channel model for frequencies from 0.5 to
  100 GHz},
  \url{http://www.3gpp.org/ftp/Specs/archive/38series/38.901/38901-e20.zip},
  Sept 2017.

\bibitem{cvxpy1}
S.~Diamond and S.~Boyd, ``{CVXPY}: {A} {P}ython-embedded modeling language for
  convex optimization,'' \emph{J. Mach. Learn. Res.}, vol.~17, no.~83, pp.
  1--5, 2016.

\end{thebibliography}
\end{document}